\documentclass[usenatbib]{mn2e}

\usepackage{amsmath}
\usepackage{graphicx}
\usepackage{natbib}
\usepackage{color}
\usepackage{xcolor}
\usepackage{times}

\usepackage{deluxetable}

\definecolor{darkgreen}{rgb}{0.05, 0.4, 0.05}
\definecolor{darkblue}{rgb}{0.1, 0.1, 0.7}

\newcommand{\snia}{SN~Ia}
\newcommand{\sneia}{SNe~Ia}

\newcommand{\chisq}{$\chi^2$}
\newcommand{\dmft}{$\Delta m_{15}(B)$}

\def\lesssim{\mathrel{\hbox{\rlap{\hbox{\lower3pt\hbox{$\sim$}}}\hbox{\raise2pt\hbox{$<$}}}}}
\def\gtrsim{\mathrel{\hbox{\rlap{\hbox{\lower3pt\hbox{$\sim$}}}\hbox{\raise2pt\hbox{$>$}}}}}

\newcommand{\nifs}{$^{56}$Ni}
\newcommand{\cofs}{$^{56}$Co}
\newcommand{\fefs}{$^{56}$Fe}
\newcommand{\coline}{[Co~{\sc iii}] $\lambda5893$}
\newcommand{\mej}{$M_{ej}$}
\newcommand{\mni}{$M_{Ni}$}
\newcommand{\mch}{$M_{Ch}$}

\newcommand{\Msol}{\ensuremath{M_\odot}}
\newcommand{\cmfgen}{{\tt CMFGEN}}

\newcommand{\ntotsne}{32}
\newcommand{\ntotspec}{94}
\newcommand{\nlitsne}{25}
\newcommand{\nlitspec}{77}
\newcommand{\nnewsne}{7}
\newcommand{\nnewspec}{17}

\newcommand{\kms}{km\,s$^{-1}$}

\newcommand{\NaI}{Na~{\sc i}}

\newcommand{\FeII}{Fe~{\sc ii}}
\newcommand{\FeIII}{Fe~{\sc iii}}
\newcommand{\CoII}{Co~{\sc ii}}
\newcommand{\CoIII}{Co~{\sc iii}}

\title[SN~Ia $^{56}$Ni masses from nebular $^{56}$Co emission]{Measuring nickel masses in Type Ia supernovae using cobalt emission in nebular phase spectra}

\author[Childress et al.]{\parbox{\textwidth}{
Michael~J.~Childress$^{1,2}$\thanks{E-mail:michael.childress@anu.edu.au},
D.~John~Hillier$^{3}$,
Ivo~Seitenzahl$^{1,2}$,
Mark~Sullivan$^{4}$,
Kate~Maguire$^{5}$,
Stefan~Taubenberger$^{5}$,
Richard~Scalzo$^{1}$,
Ashley~Ruiter$^{1,2}$,
Nadejda~Blagorodnova$^{7}$,
Yssavo~Camacho$^{8,9}$,
Jayden~Castillo$^{1}$,
Nancy~Elias-Rosa$^{10}$,
Morgan~Fraser$^{7}$,
Avishay~Gal-Yam$^{11}$,
Melissa~Graham$^{12}$,
D.~Andrew~Howell$^{13,14}$,
Cosimo~Inserra$^{15}$,
Saurabh~W.~Jha$^{9}$,
Sahana~Kumar$^{12}$,
Paolo~A.~Mazzali$^{16,17}$,
Curtis~McCully$^{13,14}$,
Antonia~Morales-Garoffolo$^{18}$,
Viraj~Pandya$^{19,9}$,
Joe~Polshaw$^{15}$,
Brian~Schmidt$^{1}$,
Stephen~Smartt$^{15}$,
Ken~W.~Smith$^{15}$,
Jesper~Sollerman$^{20}$,
Jason~Spyromilio$^{5}$,
Brad~Tucker$^{1,2}$,
Stefano~Valenti$^{13,14}$,
Nicholas~Walton$^{7}$,
Christian~Wolf$^{1}$,
Ofer~Yaron$^{11}$,
D.~R.~Young$^{15}$,
Fang~Yuan$^{1,2}$,
Bonnie~Zhang$^{1,2}$}\\ \\
\parbox{\textwidth}{
$^{1}$ Research School of Astronomy and Astrophysics, 
Australian National University, 
Canberra, ACT 2611, Australia.\\
$^{2}$ARC Centre of Excellence for All-sky Astrophysics (CAASTRO).\\
$^{3}$Department of Physics and Astronomy \& 
Pittsburgh Particle Physics, Astrophysics, and Cosmology Center (PITT PACC), 
University of Pittsburgh, 3941 O'Hara Street,
Pittsburgh, PA 15260, USA.\\
$^{4}$School of Physics and Astronomy,
University of Southampton,
Southampton, SO17 1BJ, UK.\\
$^{5}$European Organisation for Astronomical Research 
in the Southern Hemisphere (ESO), 
Karl-Schwarzschild-Str. 2, 
85748 Garching b. M\"unchen, Germany.\\
$^{7}$Institute of Astronomy, 
University of Cambridge, Madingley Rd., 
Cambridge, CB3 0HA, UK.\\
$^{8}$Department of Physics,
Lehigh University,
16 Memorial Drive East,
Bethlehem, Pennsylvania 18015, USA.\\
$^{9}$Department of Physics and Astronomy, 
Rutgers, the State University of New Jersey, 
136 Frelinghuysen Road, 
Piscataway, NJ 08854, USA.\\
$^{10}$INAF - Osservatorio Astronomico di Padova, 
vicolo dell'Osservatorio 5, 35122 Padova, Italy. \\
$^{11}$Department of Particle Physics and Astrophysics, 
The Weizmann Institute of Science, 
Rehovot 76100, Israel.\\
$^{12}$Department of Astronomy, 
University of California, 
Berkeley, CA 94720-3411, USA.\\
$^{13}$Department of Physics, 
University of California, 
Broida Hall, Mail Code 9530, 
Santa Barbara, CA 93106-9530, USA.\\
$^{14}$Las Cumbres Observatory Global Telescope Network, 
6740 Cortona Dr., Suite 102, 
Goleta, CA 93117, USA.\\
$^{15}$Astrophysics Research Centre, 
School of Mathematics and Physics, 
Queen's University Belfast, 
Belfast BT7 1NN, UK.\\
$^{16}$Astrophysics Research Institute, 
Liverpool John Moores University,
Egerton Wharf, Birkenhead, CH41 1LD, UK.\\
$^{17}$Max-Planck-Institut f\"ur Astrophysik, 
Karl-Schwarzschild str. 1, 85748 Garching, Germany.\\
$^{18}$
Institut de Ci\`encies de l'Espai (CSIC-IEEC), 
Campus UAB,  Cam\'i de Can Magrans S/N, 
08193 Cerdanyola, Spain.\\
$^{19}$Department of Astrophysical Sciences,
Princeton University,
Princeton, NJ 08544, USA.\\
$^{20}$The Oskar Klein Centre, 
Department of Astronomy, AlbaNova, 
Stockholm University, 10691 Stockholm, Sweden.\\
}
}

\begin{document}
\maketitle

\begin{abstract}
The light curves of Type Ia supernovae (\sneia) are powered by the radioactive decay of \nifs\ to \cofs\ at early times, and the decay of \cofs\ to \fefs\ from $\sim60$~days after explosion.
We examine the evolution of the \coline\ emission complex during the nebular phase for \sneia\ with multiple nebular spectra and show that the line flux follows the square of the mass of \cofs\ as a function of time.  
This result indicates both efficient local energy deposition from positrons produced in \cofs\ decay, and long-term stability of the ionization state of the nebula.
We compile \nlitspec\ nebular spectra of \nlitsne\ \snia\ from the literature and present \nnewspec\ new nebular spectra of \nnewsne\ \sneia, including SN~2014J.  
From these we measure the flux in the \coline\ line and remove its well-behaved time dependence to infer the initial mass of \nifs\ (\mni) produced in the explosion.
We then examine \nifs\ yields for different \snia\ ejected masses (\mej\ -- calculated using the relation between light curve width and ejected mass) and find the \nifs\ masses of \sneia\ fall into two regimes: for narrow light curves (low stretch $s\sim0.7$--$0.9$), \mni\ is clustered near $M_{Ni}\approx0.4M_\odot$ and shows a shallow increase as \mej\ increases from $\sim$1--1.4\Msol; at high stretch, \mej\ clusters at the Chandrasekhar mass ($1.4M_\odot$) while \mni\ spans a broad range from $0.6-1.2M_\odot$.
This could constitute evidence for two distinct \snia\ explosion mechanisms.
\end{abstract}


\section{Introduction}
\label{sec:intro}
Type Ia supernovae (\sneia) were instrumental to the discovery of the accelerating expansion of the Universe \citep{riess98, perlmutter99} and remain key tools for characterizing the precise cosmology of the Universe \citep{kessler09b, sullivan11, rest14, betoule14}.  Their cosmological utility is facilitated both by their intrinsic brightness ($M_B\sim-19$ at peak) and the relative uniformity of their peak brightnesses.  More importantly, their luminosity diversity is tightly correlated with the width of the optical light curve \citep{phillips93}.  The physical origin of this width-luminosity relation (WLR) has long been a subject of debate and is intimately tied to the progenitor system of \sneia\ and the physical mechanism that triggers the explosion.  

\sneia\ are widely believed to result from the thermonuclear disruption of a carbon-oxygen (CO) white dwarf \citep{hf60}, which has recently been supported observationally for the very nearby SN~2011fe \citep{bloom12, nugent11}.  The CO-rich material in a white dwarf (WD) is supported against gravitational collapse by electron degeneracy pressure.  A stable isolated WD lacks the internal pressure and temperature necessary to fuse CO to heavier elements \citep[but see][]{chiosi15}.  In \sneia, this balance is upset by interaction with some binary companion, which triggers runaway nuclear fusion of the CO material to heavier elements, particularly iron group elements (IGEs) dominated by radioactive \nifs.  The energy from fusion unbinds the star and ejects material at $\sim10^4$ \kms.  As the ejecta expand the decay of \nifs\ to \cofs\ (with half-life of $t_{1/2}=6.08$~days) releases energy into the ejecta which powers the optical lightcurve of the SN for the first few weeks after explosion \citep{colgate69}, including the luminous peak.  At later epochs ($t\gtrsim60$~days past peak brightness), the \snia\ lightcurve is powered by \cofs\ decay to \fefs\ (with half-life of $t_{1/2}=77.2$~days).  Thus understanding the origin of the trigger mechanism and the amount of \nifs\ produced in the explosion would reveal the critical elements that make \sneia\ such excellent cosmological tools.

The nature of the CO-WD binary companion is directly responsible for the event that triggers the \snia\ explosion.  One possible scenario is the single degenerate \citep[SD;][]{whelan73, nomoto82} scenario in which a CO-WD steadily accretes from a non-degenerate (main sequence or giant-like) companion until the central density of the WD exceeds the critical density for carbon ignition \citep[e.g.,][]{gasques05} as the mass approaches the Chandrasekhar mass ($M_{WD}\approx1.4M_\odot$).  In this scenario, the WLR has been proposed to arise from stochastic variations in the time at which the nuclear burning front within the exploding WD transitions from sub-sonic to super-sonic -- the so-called deflagration to detonation transition \citep[DDT; e.g.,][]{blinnikov86, roepke07, kasen07, kasen09, sim13}.  Variations in the time of the DDT result in different amounts of \nifs\ being produced, yielding different peak magnitudes and light curve widths for \sneia\ \citep[though][do not recover the observed WLR]{sim13}.

The other popular scenario for \snia\ progenitor systems is the double degenerate \citep[DD;][]{tutukov76, tutukov79, iben84, webbink84} scenario in which two WDs in a close binary merge after orbital decay due to gravitational radiation.  Some recent simulation results have shown that a violent merger of the two WDs produces ``hot spots'' which exceed the critical temperature and density \citep{seitenzahl09b} needed to ignite CO fusion \citep{guillochon10, loren10, pakmor10, pakmor13, moll14, raskin14}.  This scenario is inherently not tied to \mch, but instead could produce explosions with varying luminosities and light curve widths simply due to the variation in mass of the progenitor system \citep{ruiter13}.  Generally for the DD scenario, the WD undergoes a complete detonation and the amount of \nifs\ produced depends on the mass of the progenitor \citep{fink10, sim10}.

Finally, it is important to also consider the double detonation (DDet) mechanism for triggering the WD explosion.  In this scenario, helium-rich material accreted onto the surface of the white dwarf (either from a He-rich main sequence or giant star or He-WD) could ignite and send a shockwave into the core of the star.  This shock wave then triggers a second detonation near the WD core which initiates the thermonuclear runaway process \citep{livne90, iben91, ww94, fink10, wk11}.  This mechanism could arise from SD or DD systems, and is not tied to \mch.  Additionally, this mechanism may offer a favorable explanation for the presence of high-velocity features in early \snia\ spectra \citep{mazzali05b, maguire12, childress12fr, marion13, childress14, maguire14, pan15, silverman15}.

While much of the debate about \snia\ progenitors in the previous decade revolved around which single scenario was responsible for \sneia, recent results have pointed toward multiple progenitor channels being realized in nature.  \snia\ rates studies yielded evidence for both short- and long-lived progenitors \citep{mannucci05, scan05, sullivan06, mannucci06, aubourg08}.  The lack of a detected companion star to the progenitor of SN~2011fe \citep{li11fe} and in \snia\ remnants \citep[e.g.][]{schaefer12, kerzendorf12, kerzendorf13, kerzendorf14} present individual cases where the DD scenario seems necessary, while strong emission from circum-stellar material in some nearby \sneia\ \citep{hamuy03, aldering06, dilday12, silverman13a, silverman13b} seems to indicate clear cases of the SD scenario.  

For peculiar white dwarf supernovae, like the Type-Iax SN 2012Z, a luminous progenitor system has been detected and interpreted as the donor star \citep{mccully14b}. Similarly, shock interaction of SN ejecta with a (non-degenerate) companion star has been detected in the early light curve of another peculiar, low-velocity white dwarf SN \citep{cao15}.  However such shock interaction is distinctly absent for several other \sneia\ observed continuously through the epoch of first light with the Kepler satellite \citep{olling15}.  Additionally, a general dichotomy in the spectroscopic properties of \sneia\ appears evident \citep{maguire14}.  Thus numerous lines of evidence now point to multiple \snia\ progenitor channels being active.

Variations in progenitor masses between different explosion mechanisms will manifest as diversity in the bolometric light curves of \sneia\ \citep{arnett82, jeffery99, stritzinger06, roepke12}.  Recently, \citet{scalzo14a} demonstrated that the ejected mass -- hence the progenitor mass -- of a \snia\ could be recovered to 10--15\% precision, as tested on bolometric light curves derived from radiative transfer modelling of \snia\ explosion models with known input progenitor mass.  Applying the same modelling technique to real data, \citet{scalzo14a} found evidence that the ejected mass varies in the range 0.9--1.4~\Msol\ among spectroscopically normal \citep{branch93} \sneia\ and that the ejected mass also correlates strongly with the light curve width parameter used to standardize \snia\ distances in cosmology.  The correlation between ejected mass and light curve width was exploited by \citet{scalzo14c} to measure the \snia\ ejected mass distribution: they found that 25--50\% of all normal \sneia\ eject sub-Chandrasekhar masses, with most of the rest being consistent with Chandrasekhar-mass events \citep[this is consistent with constraints from Galactic chemical evolution based on Mn/Fe in the solar neighborhood --][]{seitenzahl13b}.  Super-Chandrasekhar-mass \sneia\ were found to be very rare, at most a few percent of all \sneia, consistent with previous measurements of the relative rate \citep{scalzo12}.

The diversity in ejected mass suggests a corresponding diversity in explosion mechanisms among normal \sneia.  Further information about the explosion mechanism may also be encoded in the peak absolute magnitude distribution \citep{ruiter13, ptk14}, the diversity in early \snia\ light curves \citep{dessart14a}, or in the relation between \nifs\ and ejected mass \citep{sim10, ruiter13, scalzo14a}.  The \nifs\ mass is most commonly inferred from the peak absolute magnitude of the supernova \citep{arnett82}, although with some model-dependent systematic errors \citep{branch92, hk96, howell09}. The \nifs\ mass can also be inferred from detailed modelling of photospheric phase spectral times series \citep{stehle05, mazzali08, tanaka11, sasdelli14, blondin15}.  Reliable alternative methods for measuring \nifs\ masses, with different model-dependent systematics, can thus in principle help to shed light on the explosion mechanisms and progenitor properties of \sneia.

In this work, we show that the amount of \nifs\ produced in the \snia\ explosion can be measured directly from signatures of its decay product \cofs\ in nebular phase spectra of \sneia.  Specifically, we employ the flux of the \coline\ line in spectra of \sneia\ in the nebular phase ($t\geq150$~days past maximum brightness) as a diagnostic of the mass of \cofs\ at a given epoch.
\citet{kuchner94} showed that the ratio of the flux of this line to the Fe~III line at $\lambda$4700~\AA\ as a function of SN phase followed the expected temporal evolution of the Co/Fe mass ratio, which they used as evidence for the presence of \nifs\ generated in the SN explosion.
More recently the presence of \nifs\ has been directly confirmed through $\gamma$-ray line emission from \nifs\ \citep{diehl14} and \cofs\ \citep{churazov14} lines observed by the INTEGRAL satellite for the very nearby SN~2014J.

Previous studies of \snia\ nebular spectra have collected a modest sample of spectra (a few dozen) from which important scientific results were derived.  
\citet{mazzali98} found a strong correlation between the width of nebular emission lines (specifically the \FeIII\ 4700 feature) with the SN light curve stretch, constituting evidence for greater \nifs\ production in more luminous slow-declining \sneia.  This result was combined with detailed modelling of nebular spectra (especially the 7380~\AA\ nebular line presumed to arise from stable $^{58}$Ni) to infer a common explosion mechanism for \sneia\ \citep{mazzali07}.
Nebular spectra have also been employed to place upper limits on hydrogen in the vicinity of normal \sneia\ \citep{leonard07, shappee13, bsnip5, lundqvist15}.  The lack of hydrogen in normal \sneia\ is in contrast to the strong hydrogen lines found in late phase spectra of \sneia\ which exhibited strong interaction during the photospheric phase \citep{silverman13b}.
Velocity shifts in the purported Ni 7380~\AA\ nebular line were used to infer asymmetry in the inner core of \sneia\ \citep{maeda10a, maeda10b}, which was also found to correlate with the optical colour and Si 6355~\AA\ velocity gradient during the photospheric phase \citep{maeda11}.  These line velocity shifts were found to also correlate with photospheric phase spectropolarimetry \citep{maund10}, indicating a general correlated asymmetric geometry for \sneia.
These early results have generally been supported with greater statistics afforded by new large data sets such as the CfA sample \citep{blondin12} and BSNIP \citep{bsnip5}.

Until recently, the nebular line at 5890~\AA\ was not frequently emphasized as a diagnostic of \cofs\ due to its presumed association with emission from sodium \citep[][are noteworthy exceptions]{kuchner94, mcclelland13}.  However \citet{dessart14b} showed definitively that this line arises primarily from cobalt for the majority of \sneia.
We exploit this result to use the \coline\ line as a diagnostic of \nifs\ from a large sample of nebular \snia\ spectra compiled from both new observations and from the literature.
Equipped with a sample of \nlitspec\ spectra of \nlitsne\ \sneia\ from the literature and \nnewspec\ new spectra of \nnewsne\ \sneia, we calculate the {\em absolute} flux of the nebular \coline\ line by scaling the spectra to flux-calibrated photometry measurements.  With these calibrated fluxes we show that the temporal evolution of the absolute \coline\ line flux is highly consistent for \sneia\ with multiple nebular spectra.  We exploit this result to place measurements from disparate epochs on a common scale.  This allows us to meaningfully compare the line fluxes in order to determine the relative amount of \nifs\ produced by each \snia\ in our sample.

In Section~\ref{sec:data} we present our compilation of literature \snia\ spectra and the new nebular \snia\ data released here.  Section~\ref{sec:neb_line_fluxes} presents our method for measuring the \coline\ flux from the spectra and scaling the spectra with the \snia\ photometry.  We examine the temporal evolution of the \coline\ for \sneia\ with numerous nebular observations in Section~\ref{sec:neb_time_series}.  We then infer \nifs\ masses for our \snia\ sample in Section~\ref{sec:nickel_mass}, and discuss the implications and limitations of our results in Section~\ref{sec:discussion}.  Finally we conclude in Section~\ref{sec:conclusions}.

\section{\snia\ Nebular Spectroscopy Data}
\label{sec:data}
The analysis in this work relies on a compilation of \snia\ nebular spectra from the literature as well as new observations.  The full sample of literature and new late phase spectra are presented in Table~\ref{tab:lit_neb_spectra}.

\begin{deluxetable}{lrcl}
\tablewidth{0pt}
\tablecaption{New and Literature Late Phase SN Ia Spectra
\label{tab:lit_neb_spectra}}
\tabletypesize{\small}
\tablehead{
\colhead{SN} & 
\colhead{Phase $t$~$^a$} & 
\colhead{Obs. Date $^b$} & 
\colhead{Spec. Ref. $^c$} \\
\colhead{} & 
\colhead{(days)} & 
\colhead{} &
}
\startdata
SN~1990N   &  160  &      19901217  &  BSNIP  \\
           &  186  &      19910112  &  \citet{gomez98}  \\
           &  227  &      19910222  &  \citet{gomez98}  \\
           &  255  &      19910322  &  \citet{gomez98}  \\
           &  280  &      19910416  &  \citet{gomez98}  \\
           &  333  &      19910608  &  \citet{gomez98}  \\
SN~1991T   &  113  & {\em 19910819} &  BSNIP  \\
           &  186  &      19911031  &  BSNIP  \\
           &  258  &      19920111  &  \citet{gomez98}  \\
           &  320  &      19920313  &  BSNIP  \\
           &  349  &      19920411  &  BSNIP  \\
SN~1994ae  &  144  &      19950422  &  BSNIP  \\
           &  153  &      19950501  &  CfA  \\
SN~1995D   &  277  &      19951124  &  CfA  \\
           &  285  &      19951202  &  CfA  \\
SN~1998aq  &  211  &      19981124  &  \citet{branch03}  \\
           &  231  &      19981214  &  \citet{branch03}  \\
           &  241  &      19981224  &  \citet{branch03}  \\
SN~1998bu  &  179  &      19981114  &  CfA  \\
           &  190  &      19981125  &  CfA  \\
           &  208  &      19981213  &  CfA  \\
           &  217  &      19981222  &  CfA  \\
           &  236  &      19990110  &  BSNIP  \\
           &  243  &      19990117  &  CfA  \\
           &  280  &      19990223  &  BSNIP  \\
           &  329  &      19990413  &  \citet{cappellaro01}  \\
           &  340  &      19990424  &  BSNIP  \\
SN~1999aa  &  256  &      19991109  &  BSNIP  \\
           &  282  &      19991205  &  BSNIP  \\
SN~2002cs  &  174  &      20021106  &  BSNIP  \\
SN~2002dj  &  222  &      20030201  &  \citet{pignata08}  \\
           &  275  &      20030326  &  \citet{pignata08}  \\
SN~2002er  &  216  &      20030410  &  \citet{kotak05}  \\
SN~2002fk  &  150  &      20030227  &  BSNIP  \\
SN~2003du  &  109  & {\em 20030823} &  \citet{stanishev07}  \\
           &  138  & {\em 20030921} &  \citet{anupama05}  \\
           &  139  & {\em 20030922} &  \citet{anupama05}  \\
           &  142  & {\em 20030925} &  \citet{stanishev07}  \\
           &  209  &      20031201  &  \citet{stanishev07}  \\
           &  221  &      20031213  &  \citet{stanishev07}  \\
           &  272  &      20040202  &  \citet{stanishev07}  \\
           &  377  &      20040517  &  \citet{stanishev07}  \\
SN~2003hv  &  113  & {\em 20031228} &  \citet{leloudas09}  \\
           &  145  & {\em 20040129} &  \citet{leloudas09}  \\
           &  323  &      20040725  &  \citet{leloudas09}  \\
SN~2004bv  &  171  &      20041114  &  BSNIP  \\
SN~2004eo  &  228  &      20050516  &  \citet{pastorello07}  \\
SN~2005cf  &  319  &      20060427  &  \citet{wang05cf}  \\
SN~2007af  &  103  & {\em 20070620} &  CfA  \\
           &  108  & {\em 20070625} &  CfA  \\
           &  120  & {\em 20070707} &  BSNIP  \\
           &  123  & {\em 20070710} &  CfA  \\
           &  128  & {\em 20070715} &  BSNIP  \\
           &  131  & {\em 20070718} &  CfA  \\
           &  151  &      20070807  &  BSNIP  \\
           &  165  &      20070821  &  BSNIP  \\
           &  308  &      20080111  &  CfA  \\
SN~2007gi  &  161  &      20080115  &  \citet{zhang10}  \\
SN~2007le  &  317  &      20080827  &  BSNIP  \\
SN~2007sr  &  177  &      20080623  &  CfA  \\
SN~2009le  &  324  &      20101016  &  T15b \\
SN~2011by  &  206  &      20111202  &  \citet{bsnip5}  \\
           &  310  &      20120315  &  \citet{bsnip5}  \\
SN~2011fe  &   74  & {\em 20111123} &  \citet{shappee13}  \\
           &  114  & {\em 20120102} &  \citet{shappee13}  \\
           &  196  &      20120324  &  \citet{shappee13}  \\
           &  230  &      20120427  &  \citet{shappee13}  \\
           &  276  &      20120612  &  \citet{shappee13}  \\
           &  314  &      20120720  &  \citet{tauben15}  \\
SN~2011iv  &  318  &      20121024  &  T15b \\
SN~2012cg  &  330  &      20130507  &  M15 \\
           &  342  &      20130513  &  T15b \\
SN~2012fr  &  101  & {\em 20130221} &  This work \\
           &  116  & {\em 20130308} &  This work \\
           &  125  & {\em 20130317} &  This work \\
           &  151  &      20130412  &  This work \\
           &  222  &      20130622  &  This work \\
           &  261  &      20130731  &  This work \\
           &  340  &      20131018  &  This work \\
           &  357  &      20131103  &  M15 \\
           &  367  &      20131114  &  This work \\
SN~2012hr  &  283  &      20131006  &  This work \\
           &  368  &      20131230  &  This work \\
SN~2013aa  &  137  & {\em 20130710} &  This work \\
           &  185  &      20130827  &  This work \\
           &  202  &      20130913  &  This work \\
           &  342  &      20140131  &  This work \\
           &  358  &      20140216  &  M15 \\
           &  430  &      20140422  &  M15 \\
SN~2013cs  &  320  &      20140322  &  This work \\
SN~2013dy  &  333  &      20140626  &  \citet{pan15} \\
           &  419  &      20140920  &  This work \\
SN~2013gy  &  276  &      20140920  &  This work \\
SN~2014J   &  231  &      20140920  &  This work \\
\enddata
\tablecomments{
$^a$ With respect to date of $B$-band peak brightness.\\
$^b$ Observation dates that are {\em italicized} are not used to measure \mni, and are only employed in Section~\ref{sec:neb_time_series}\\
$^c$ BSNIP: \citet{bsnip1}; CfA: \citet{matheson08, blondin12}; M15: Maguire et al. (2015), in preparation; T15b: Taubenberger et al. (2015b), in preparation.
}
\end{deluxetable}

\subsection{Compilation of Literature Data}
\label{sec:lit_data}
For reasons outlined in Section~\ref{sec:neb_time_series}, the earliest epochs from which we can use \coline\ line fluxes is at phase $t=+150$~days.  In practice, we found for most spectra beyond $t\approx+400$~days that the \coline\ flux was too weak to be usable for our preferred analysis.  Furthermore, \citet{tauben15} recently showed that the $t=+1000$~day spectrum of SN~2011fe showed dramatic changes in its structure, likely arising from a change in the ionization condition of the nebula.  Indeed, this ionization change appears evident in the $t=+590$~day spectrum presented in \citet{graham15b}, and we see evidence for the onset of this change shortly after $t\approx+400$~days in the data gathered for this analysis.  Thus we excise data later than $t\approx+400$~days as unreliable due to low signal and likely ionization change (we examine potential impact from the latter effect in Section~\ref{sec:nebular_stability}).

To begin compiling a sample that meets these phase criteria, we performed a large query of the WISeREP\footnote{http://wiserep.weizmann.ac.il} \citep{wiserep} database to search for \sneia\ with two spectroscopic observations separated by at least 100 days -- assuming the earlier one would be near maximum light, this singles out \sneia\ with nebular spectra.
We then require SNe to have photospheric-phase optical light curves sufficient to robustly establish light curve stretch, colour, and the date of maximum light using SiFTO \citep{sifto}.
We also require the spectra to have sufficiently high signal-to-noise so that the \coline\ line can be well fit using a Gaussian fitting procedure (see Section~\ref{sec:neb_line_fluxes}).
SN~2006X was excluded (despite having numerous nebular spectra) due to significant variability in its sodium features \citep{patat07} and a rather significant light echo \citep{wang08b, crotts08}, both of which might affect the time evolution of the \coline\ flux.

Finally, we excise any \sneia\ which are spectroscopically peculiar in the nebular phase: \sneia\ similar to SN~1991bg \citep{filippenko91bg, leibundgut91bg} exhibit extremely narrow Fe lines and unusual line ratios; Ia-CSM SNe \citep{silverman13a} are excluded due to possible impact of CSM on the nebular emission; SNe~Iax \citep{foley13} are excised as these probably arise from a different physical mechanism than normal \sneia; candidate super-Chandrasekhar \sneia\ \citep{howell06} are excised due to their unusual nebular spectra \citep{tauben13}.  \sneia\ similar to SN~1991T \citep{phillips91T, filippenko91T} or SN~1999aa {\em are} however included in the sample, as their ionization structure appears to be similar to ``normal'' \sneia.

In summary, the selection criteria for our sample of literature nebular \snia\ spectra are:
\begin{itemize}
  \item Phase (with respect to $B$-band maximum light) in the range $+150 \leq t \leq +400$
  \item Well-sampled multi-colour photospheric phase light curve (such that the light curve fitter SiFTO converges)
  \item Sufficient spectrum S/N to measure the \coline\ line center and width
  \item No spectroscopic peculiarity, except SN~1991T-like
\end{itemize}
The full sample of spectra which meet these criteria are presented in Table~\ref{tab:lit_neb_spectra}, and comprise \nlitspec\ spectra of \nlitsne\ \sneia\ from the literature.

Finally we note that two of the SNe in our sample had prominent light echoes at late times: SN~1991T \citep{schmidt94} and SN~1998bu \citep{spyromilio04}.  For both of these SNe, the light echo contributions are negligible at the spectroscopic epochs we employ.

\subsection{New \snia\ Nebular Spectroscopy}
\label{sec:new_data}
\begin{figure*}
\begin{center}
\includegraphics[width=0.95\textwidth]{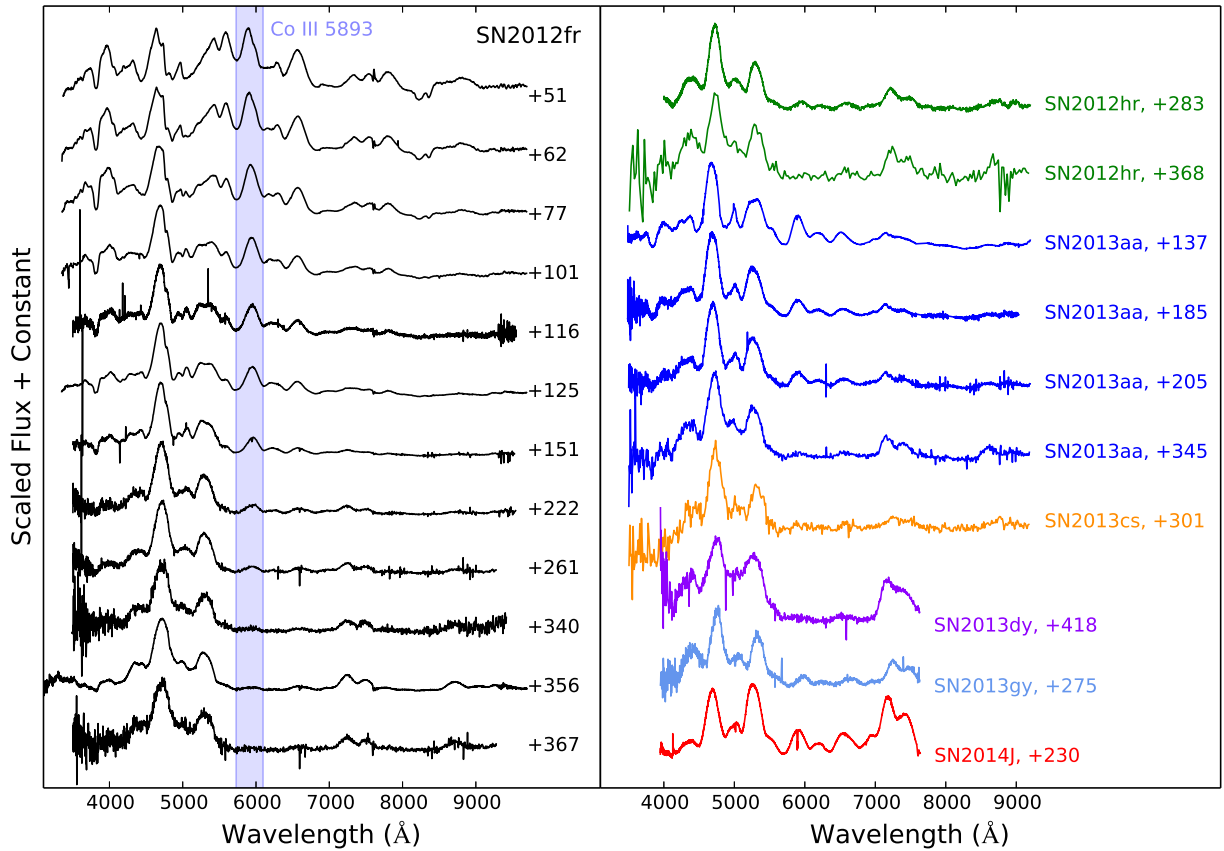}
\caption{New late phase and nebular spectra of \sneia\ presented in this work.  All spectra are publicly available on WISeREP (except the +356~day spectrum of SN~2012fr from K15).  Some spectra have been slightly binned (to $\sim5$~\AA) for visual clarity.}
\label{fig:new_neb_spectra}
\end{center}
\end{figure*}

We obtained new late phase ($+50 \leq t \leq +150$~days) and nebular ($t \geq +150$~days) spectra of \nnewsne\ nearby \sneia\ from numerous telescopes.  These spectra have been released publicly on WISeREP, with several spectra of SN~2012fr released through PESSTO's ESO data releases \footnote{www.pessto.org}.  Information about observation details are presented in Table~\ref{tab:new_neb_spectra} and a plot of the spectra is shown in Figure~\ref{fig:new_neb_spectra}.  We note these spectra have not been rescaled to match observed photometry.

Several late phase spectra of very nearby \sneia\ were collected with the Wide Field Spectrograph \citep[WiFeS;][]{dopita07, dopita10} on the ANU 2.3m telescope at Siding Spring Observatory in northern New South Wales, Australia.  Observations were performed with the B3000 and R3000 gratings with the RT560 dichroic, giving wavelength range of 3500~\AA-9800~\AA, with resolution of 0.8~\AA\ and 1.2~\AA\ on the blue and red arms, respectively.  Data were reduced using the PyWiFeS package \citep{pywifes}, and spectra were extracted using our custom GUI \citep[see e.g.,][]{childress12fr}.  We generally observed during very dark nights (moon illumination less than 20\%) when the seeing was favorable (1\farcs5-2\farcs0).  We note that the WiFeS spectra of SN~2012hr and SN~2013cs have too low signal-to-noise to obtain a reliable measurement of the \coline\ line flux, but we release them publicly (on WISeREP) here.

New nebular spectra for three nearby \sneia\ were collected with DEIMOS \citep{deimos} on the Keck-II telescope on Mauna Kea, Hawaii.  Observations were conducted with a 1\farcs5 longslit, the 600 l/mm grating with a central wavelength of 5000~\AA\ and with the GG410 order blocking filter, yielding a wavelength range of 4000~\AA-7650~\AA\ with 0.6~\AA\ resolution.  Data were reduced using standard techniques in IRAF \citep[see e.g.,][]{childress13a}, with the blue and red chips reduced separately then combined as a final step.  We employed the Mauna Kea extinction curve of \citet{buton13}.  Our observations come from a single night on Keck (2014-Sep-20 UTC) when conditions were less favorable (high humidity and thick clouds, ultimately 50\% time lost to weather) but with a median seeing of 0\farcs9.

Five additional late phase spectra of SN~2012fr were collected as part of the Public ESO Spectroscopic Survey of Transient Objects \citep[PESSTO;][]{pessto} during early 2013, and reduced with the PESSTO pipeline as described in \citet{pessto}.  One spectrum of SN~2012fr and two spectra of SN~2013aa were obtained in 2013 using the Robert Stobie Spectrograph on the South African Large Telescope (SALT), and reduced using a custom pipeline that incorporates PyRAF and PySALT \citep{crawford10}.  One spectrum of SN~2012hr was obtained with Gemini GMOS \citep{gmos} using the 0.75\arcsec\ longslit with the B600 and R400 gratings in sequence to yield a spectral coverage from 4000 -- 9600 \AA, under program GS-2013B-Q-48 (PI: Graham) -- the spectrum was reduced using the Gemini IRAF package.

In the analysis below we also include nebular spectroscopy samples from forthcoming analyses by Maguire et al. (2015, in prep. -- hereafter M15) and Taubenberger et al. (2015b, in prep. -- hereafter T15b).  The M15 sample were obtained over a multi-period program at the VLT using XShooter \citep{xshooter}, and were reduced with the XShooter pipeline \citep{xshredux} using standard procedures \citep[as in][]{maguire13}.  The T15b sample were observed as part of a separate multi-period program using FORS2 on the VLT, and data were reduced with standard procedures similar to those employed in \citet{tauben13}.

\begin{table}
\begin{center}
\caption{Observation details for new late phase SN Ia spectra}
\label{tab:new_neb_spectra}
\begin{tabular}{lrcc}
\hline
SN      & Phase  & Obs. & Telescope    \\
        & (days) & Date & / Instrument \\
\hline
SN~2012fr &  +51 & 2013-Jan-02  & NTT-3.6m / EFOSC \\
          &  +62 & 2013-Jan-13  & NTT-3.6m / EFOSC \\
          &  +77 & 2013-Jan-28  & NTT-3.6m / EFOSC \\
          & +101 & 2013-Feb-21  & NTT-3.6m / EFOSC \\
          & +116 & 2013-Mar-08  & ANU-2.3m / WiFeS \\
          & +125 & 2013-Mar-17  & NTT-3.6m / EFOSC \\
          & +151 & 2013-Apr-12  & ANU-2.3m / WiFeS \\
          & +222 & 2013-Jun-22  & ANU-2.3m / WiFeS \\
          & +261 & 2013-Jul-31  & ANU-2.3m / WiFeS \\
          & +340 & 2013-Oct-18  & SALT / RSS \\
          & +367 & 2013-Nov-14  & ANU-2.3m / WiFeS \\
SN~2012hr & +283 & 2013-Oct-06  & Gemini / GMOS \\
          & +368 & 2013-Dec-30  & ANU-2.3m / WiFeS \\
SN~2013aa & +137 & 2013-Jul-10  & SALT / RSS \\
          & +185 & 2013-Aug-27  & SALT / RSS \\
          & +202 & 2013-Sep-13  & ANU-2.3m / WiFeS \\
          & +342 & 2014-Jan-31  & ANU-2.3m / WiFeS \\
SN~2013cs & +320 & 2014-Mar-22  & ANU-2.3m / WiFeS \\
SN~2013dy & +419 & 2014-Sep-20  & Keck-II / DEIMOS \\
SN~2013gy & +276 & 2014-Sep-20  & Keck-II / DEIMOS \\
SN~2014J  & +231 & 2014-Sep-20  & Keck-II / DEIMOS \\
\hline
\end{tabular}
\end{center}
\end{table}

\section{Nebular line flux measurements}
\label{sec:neb_line_fluxes}

\subsection{The \coline\ line in the nebular phase: a radiative transfer perspective}
\label{sec:line_theory}
The current study was motivated by the disappearance of \CoIII\ lines in nebular time series spectra, most notably the feature near 5900~\AA.  Previous literature analyses have attributed this feature alternately to \CoIII\ and \NaI, so we turned to radiative transfer calculations to settle this ambiguity.

\begin{figure*}
\begin{center}
\includegraphics[width=0.49\textwidth]{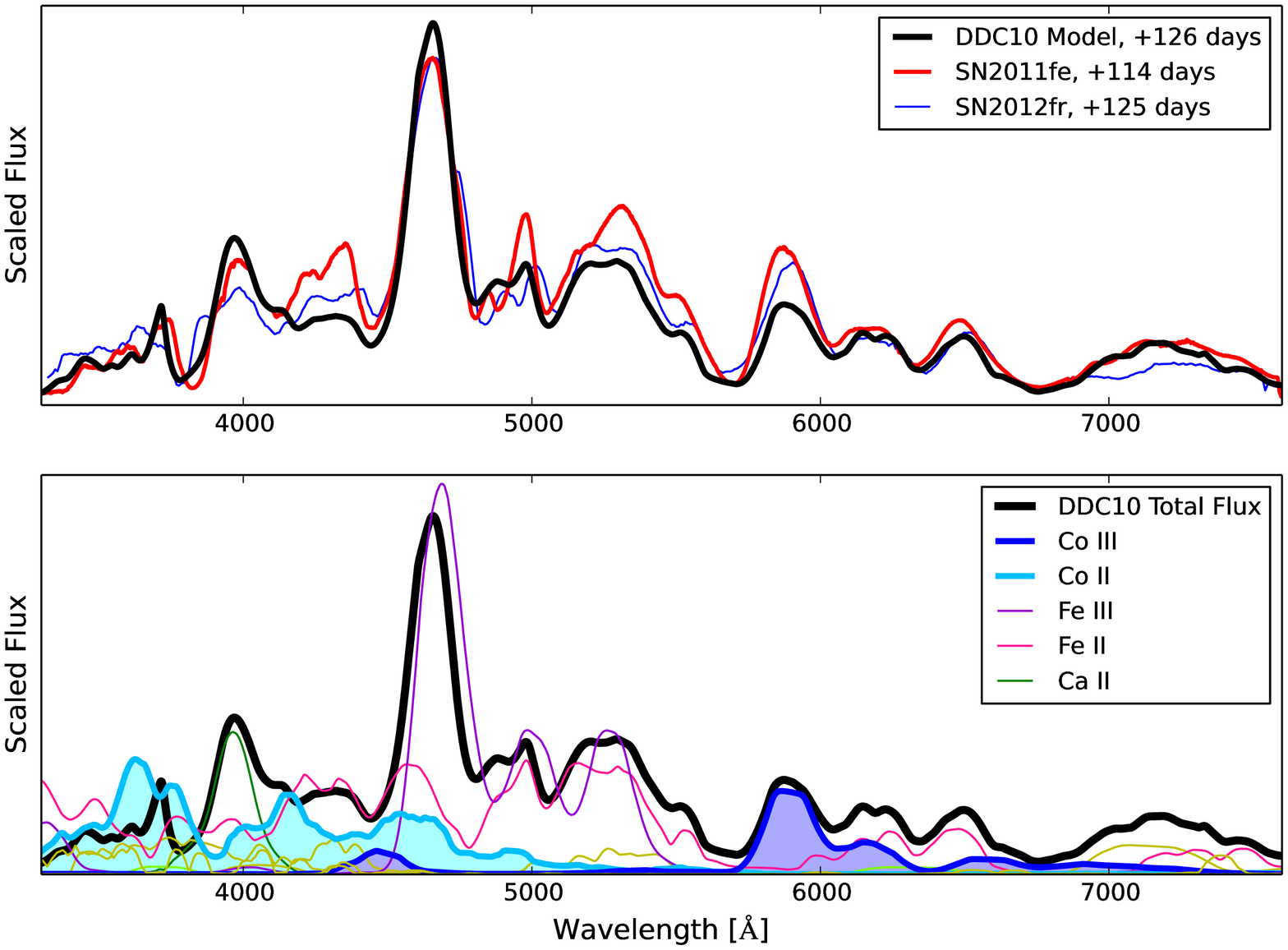}
\includegraphics[width=0.49\textwidth]{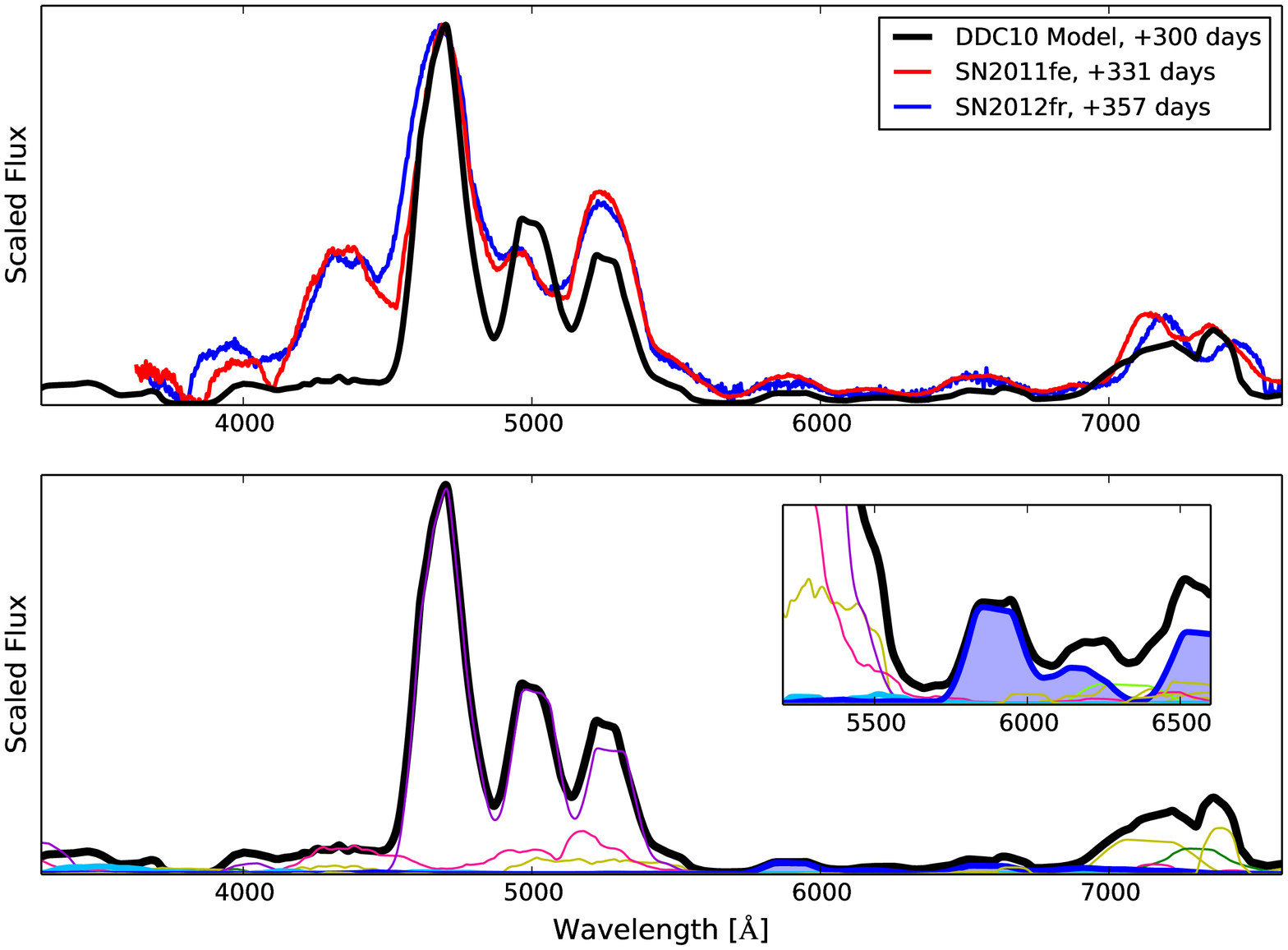}
\caption{Top panels: Comparison of radiative transfer (\cmfgen) model spectrum for the DDC10 \citep{blondin13} delayed detonation model at very late epochs (left: +126~days, right: +300~days) compared to contemporaneous data for SN~2011fe and SN~2012fr.  Bottom panels: Emission spectra for various ions from \cmfgen\ for late-phase DDC10 models (epochs as above).}
\label{fig:cmfgen_models}
\end{center}
\end{figure*}

We employed the time-dependent radiative transfer code \cmfgen\ \citep{cmfgen}, which solves the time dependent radiative transfer equation simultaneously with the kinetic equations. Given an initial explosion model, we self-consistently solve for the temperature structure, the ionization structure, and the non-LTE populations, beginning the calculations at 0.5 to 1 day after the explosion. The modelling assumes homologous expansion, typically uses a 10\% time step, and no changes are made to the ejecta structure (other than that required by homologous expansion) as the ejecta evolve in time.  Further details about the general model set up, and model atoms, can be found in \citet{dessart14a}.  We deployed \cmfgen\ on a delayed-detonation model \citep[DDC10 --][]{blondin13, dessart14a} at very late phases and examine the contribution of various ions to the nebular emission spectrum.  Radiative transfer calculations for this model, and similar models but with a different initial \nifs\ mass, have shown favorable agreement with observations \citep{blondin13, blondin15, dessart14a, dessart14b, dessart14c}.

In Figure~\ref{fig:cmfgen_models} we show DDC10 modeled with \cmfgen\ at phases +126~days (left panels) and +300~days (right panels).  The top panels in each column show the integrated DDC10 model flux compared to observations of nebular phase \sneia\ at similar phases, while the bottom panels show the line emission from individual ions (note this can exceed the integrated flux due to the net opacity encountered by photons following their initial emission).  The +126~day model shows particularly good agreement with the data.  At +300~days the model shows some discrepancy with the data, particularly in the ionization state of the nebula.

Most importantly, the radiative transfer calculations show that the emission feature near 5900~\AA\ is clearly dominated by \CoIII\ emission, with little or no contamination from other species.  Few other features in the optical region of the spectrum show such clean association with a single ion.

For later aspects of our analysis we require the velocity center of the nebula, which we calculate from the \coline\ line.  To do so requires an accurate calculation of the mean rest wavelength for this line complex.  The \coline\ arises from the 3d$^7$ a$^4$F-- 3d$^7$ a$^2$G multiplet, and is actually a blend of two lines --- one at 5888.5\,\AA\ and a second, but weaker, line at 5906.8\,\AA\ (see Appendix~\ref{app:co_atomic_data} and Table~\ref{tab:co_atomic_data}).  Given the $A$ values and wavelengths of the transitions contributing to the line complex, the weighted mean rest wavelength of the \CoIII\ line is 5892.7~\AA\ (note: this and previous are air wavelengths).  Henceforth we use this value for calculating line velocities.

\subsection{Measuring the \coline\ line flux}
\label{sec:line_measure}
For the main analyses in this work we focus on the flux in the \coline\ line.  We measure the flux in this line as follows.

\begin{figure}
\begin{center}
\includegraphics[width=0.45\textwidth]{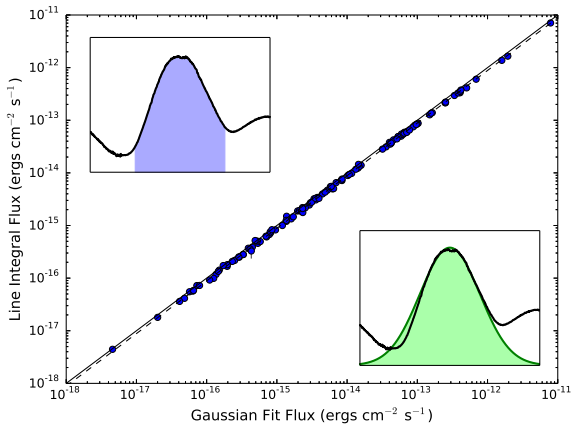}
\caption{Comparison of flux in the \coline\ line measured in two ways: strict integration of the spectrum flux within $\pm1.5\sigma$ of the fitted line center (y-axis and upper left inset), and the formal integral of the best fit Gaussian profile (x-axis and lower right inset).  The solid line represents unity, while the dashed line is the mean ratio of the integral flux to Gaussian flux for the full sample ($0.87 \pm 0.05$).}
\label{fig:line_flux_measure}
\end{center}
\end{figure}

We perform an initial Gaussian fit to the \coline\ line in order to the determine the center and width of the line.  We then integrate the flux within $\pm1.5\sigma$ of the fitted line center and use this ``integral'' flux for the remainder of this paper.  This integral boundary was chosen as a compromise between capturing a large fraction of the emitted line flux (97\% for a strictly Gaussian profile) and limiting contamination from neighbouring emission lines.  For \sneia\ with multiple nebular spectra we enforce common wavelength bounds for the flux integration at all epochs, as determined by the median fitted line center and width values across all epochs.  Generally the integrated line flux and that calculated from the best Gaussian fit showed excellent agreement (see Figure~\ref{fig:line_flux_measure}), but we prefer the integral flux as this is robust against non-Gaussianity of the line profile.

To place our \coline\ line flux measurements on the correct absolute scale, we must ensure the spectra have the correct {\em absolute} flux calibration.  To achieve this, we measure the expected $B$-band flux in the spectrum by convolving it with the $B$-band filter throughput curve and integrating.  We then compute the ratio of this flux collected in the spectrum $B$ passband to the true $B$-band flux of the SN at that epoch.  The latter is determined from the late-time photometry for each of our SNe, as outlined in Appendix~\ref{app:snia_data} and presented in Table~\ref{tab:late_phot}.  To ensure reproducability of our results, we report in this table the flux values derived from the raw measurements made from the spectra in their published form.

We note that normalization with the $B$-band could introduce errors in the \coline\ flux due to chromatic errors in the spectrum's flux calibration.  However, previous authors consistently performed chromatic flux calibration using spectrophotometric standard stars, typically yielding excellent colour agreement with observed photometry (e.g. $B-V$ scatter of 0.08~mag and 0.10 mag for the CfA and BSNIP samples, repectively).  We also note that other systematic effects could affect our measurements of \coline\ line flux.  These include contamination from neighboring nebular emission lines (e.g. \FeII\ lines, see Figure~\ref{fig:cmfgen_models}), residual host galaxy light, or perhaps even previously undetected light echoes \citep[see, e.g.,][]{spyromilio04}.  Thus we expect a conservative estimate for the systematic uncertainty in the \coline\ flux measurement to be about 10\% of the measured flux.

The final integrated \coline\ line flux, wavelength bounds for the integral, and synthetic $B$-band flux integrated from the spectrum are all presented in Table~\ref{tab:line_fit_results}.  Variance spectra were not available for many of the literature \snia\ spectra in our analysis.  To correct this, we smooth the spectrum with a Savitszky-Golay filter \citep{sg64}, then smooth the squared residuals of the data from this smooth curve to derive a variance spectrum measured directly from the noise in the data \citep[as we did for data in][]{childress14}.  \coline\ line flux errors were then determined from these corrected variance spectra.

\section{Evolution of the \coline\ line flux}
\label{sec:neb_time_series}
%
\subsection{Theoretical expectations for \coline\ evolution}
\label{sec:co_evo_theory}
The decay of \cofs\ to \fefs\ produces positrons and energetic gamma-rays.  The charged positrons carry kinetic energy which they lose to the surrounding medium via Coulomb interactions.  At the nebular densities present at late times, the length scale for positron energy deposition is much smaller than the size of the nebula so the positrons essentially deposit all of their kinetic energy locally \citep{chan93}.  gamma-rays -- either those emitted directly from \cofs\ decay or created when the positrons annihilate -- are subject to radiative transfer effects and will eventually free stream as the SN nebula expands and decreases its density enough to become optically thin to gamma-rays.  The onset of this phase -- where positrons deposit a constant fraction of energy into the SN nebula and gamma-rays escape fully -- has been observed in late \snia\ bolometric light curves \citep[e.g.][]{sollerman04, ss07, leloudas09, kerzendorf14c}.

Our expectation from a simple energetics perspective is that the flux of the \coline\ line should evolve as the square of the mass of cobalt as a function of time $M_{Co}(t)$.  The energy being deposited into the nebula at these late phases arises from the positrons produced in \cofs\ decay, and thus should scale with the mass of cobalt.  If this energy is evenly deposited amongst all species in the nebula then the fraction of that energy absorbed by the cobalt atoms should be proportional to the mass fraction of cobalt.  Thus the amount of energy absorbed by cobalt atoms follows the square of the cobalt mass as a function of time.  If the fraction of that energy emitted in the \coline\ line remains constant (see Section~\ref{sec:nebular_stability}) then we expect a net quadratic dependence of the \coline\ line luminosity on the mass of cobalt as a funciton of time.

Observational evidence for this temporal evolution of the \coline\ line should be expected from prior results.  The late-phase bolometric light curves of \sneia\ closely follow the amount of energy deposited by the decay of \cofs\ \citep[see, e.g.,][]{sollerman04}.  It was also demonstrated by \citet{kuchner94} that the ratio of \coline\ to Fe 4700 emission follows the Co/Fe mass ratio (as noted above), and the Fe 4700 line flux generally scales with the total luminosity of the SN since Fe is the primary coolant.  These facts combine to lend an expectation that the net emission from the \coline\ line should scale quadratically with the mass of Co in the SN nebula as a function of time.  Indeed, \citet{mcclelland13} found such a quadratic dependence for the \coline\ line in SN~2011fe.

The above reasoning for $M_{Co}^2$ dependence of the \coline\ flux holds for epochs when the nebula is fully transparent to gamma-rays.  Thus it is important to inspect the theoretical expectation for the timing of this gamma-ray transparency in the IGE zone.  The energy released per decay from \cofs\ is 3.525\,MeV, of which 3.3\% is associated with the kinetic energy of the positrons, and we have ignored the energy associated with neutrinos. As the expansion is homologous, the optical depth associated with gamma-rays scales as $1/t^2$. Assuming that the kinetic energy of the positrons is captured locally, the energy absorbed per \cofs\ decay in MeV is
\begin{equation}
  e_{Co} = 0.116 + 3.409 \left( 1 - \exp[-\tau_o (t_o/t)^2] \right)
\label{eq:co_decay_energy}
\end{equation}
where $\tau_o$ is the effective optical depth at a time $t_o$.  
%
If we denote $t_c$ as the time at which energy deposition by gamma-rays and positrons are equal, then Equation~\ref{eq:co_decay_energy} can be rewritten as:
\begin{equation}
  E_{Co} \propto M_{Co} \left(1-0.967\exp\left[-0.0346(t_c/t)^2\right]\right)
\label{eq:co_energy_deposition}
\end{equation}
We expect the flux from the \coline\ line would further scale as:
\begin{equation}
  F_{Co} \propto E_{Co} \times \frac{ xM_{Co}/M_{IGE}}{1+(a-1)M_{Co}/M_{IGE} + bM_{Ot}/M_{IGE}}
\label{eq:co_flux_scaling}
\end{equation}
where $M_{IGE}$ is the total mass of the IGE zone, $a$ and $b$ are respectively the (time-dependent) factors relating the cooling efficiency of Co and other species (which have total mass of $M_{Ot}$) relative to iron, and $x$ is the factor scaling the emission in the \coline\ feature.  If the thermal conditions in the SN nebula are relatively stable (i.e. constant $x$) and cooling by non-iron species is negligible (i.e. the above denominator goes to unity), then the line flux simply becomes proportional to $M_{Co}/M_{IGE}$.  Combining Equations \ref{eq:co_energy_deposition} and \ref{eq:co_flux_scaling} yields:
\begin{equation}
  F_{Co} \propto M_{Co}^2\left(1-0.967\exp\left[-0.0346(t_c/t)^2\right]\right)
\label{eq:final_co_scaling}
\end{equation}
For the DDC10 model, we find $t_c\sim214$~days (from explosion) -- this would imply a deviation from $M_{Co}^2$ of a factor of 2 from +150 to +400 days past maximum light (assuming a rise time of $\sim17$~days), or a factor of 1.5 from +200 to +400 days.  Alternatively if $t_c\sim80$~days (see Section~\ref{sec:co_evo_gamma}) then the deviation from $M_{Co}^2$ is only 20\% from +150 to +400 days and 10\% from +200 to +400 days.

\subsection{Observed \coline\ evolution in nebular spectral time series}
\label{sec:co_evo_obs}
To examine the observed evolution of the \coline\ line, we turn to those \sneia\ with numerous nebular spectra.  Specifically, we isolate the subset of \sneia\ in our sample with at least three epochs of observation later than +150~days past maximum.  For the eight \sneia\ in our sample which meet this criterion, we also collect spectra between $+100 \leq t \leq +150$ days past maximum (dates listed in italics in Table~\ref{tab:lit_neb_spectra}). These additional spectra allow us to further inspect the \coline\ flux evolution, but these spectra are not employed in our nickel mass estimates derived in Section~\ref{sec:nickel_mass}.

\begin{figure}
\begin{center}
\includegraphics[width=0.45\textwidth]{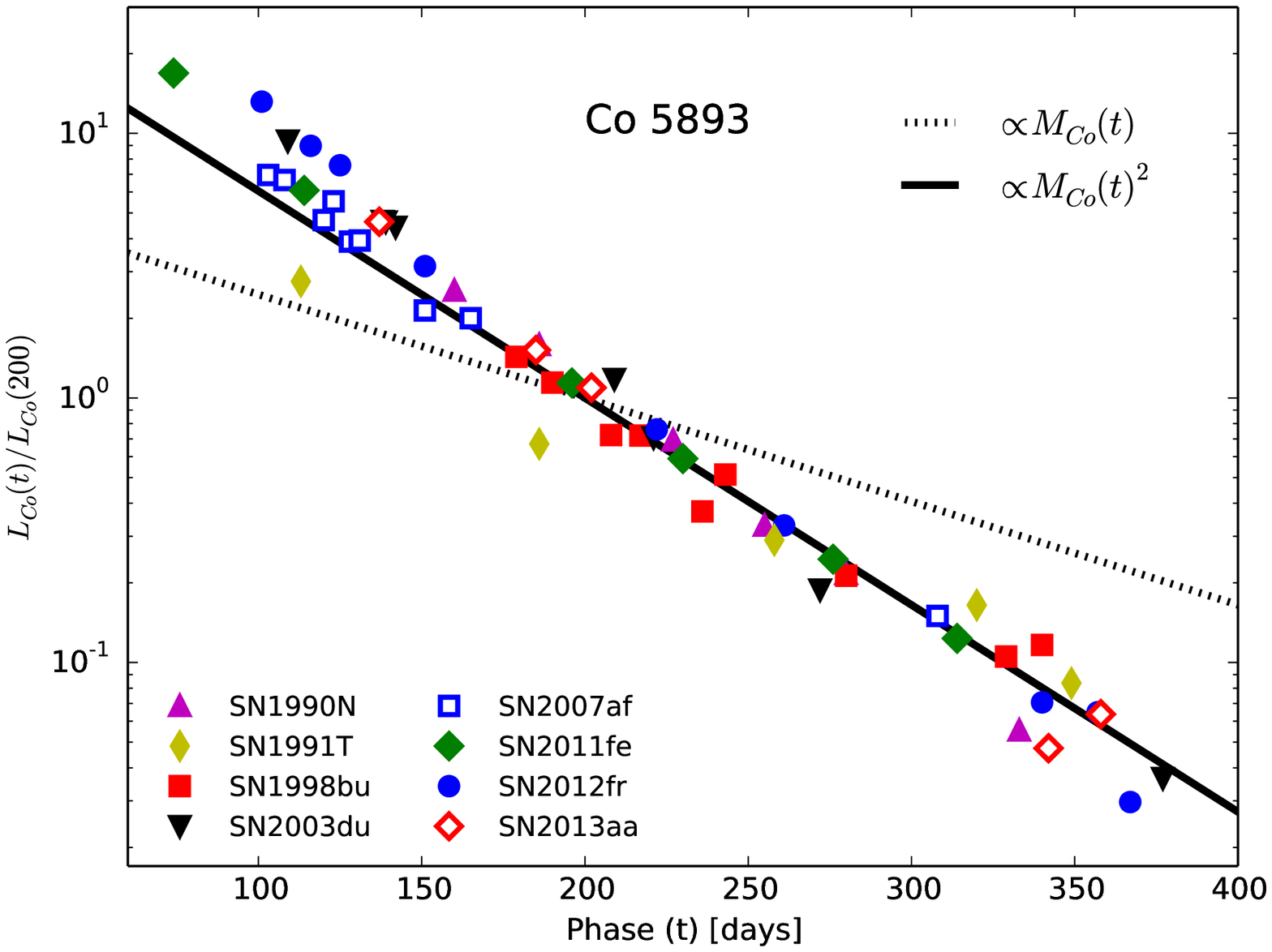}
\includegraphics[width=0.45\textwidth]{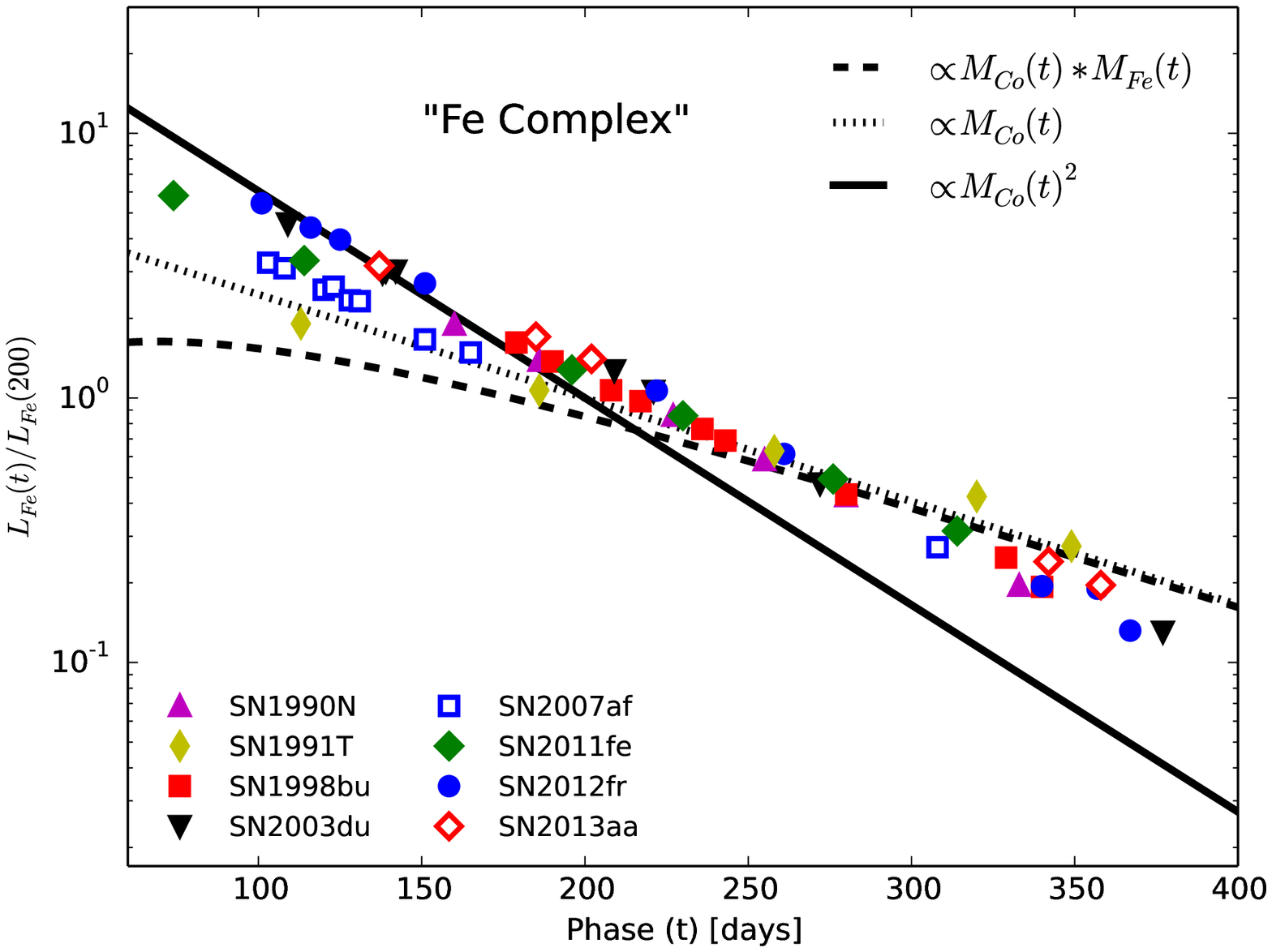}
\caption{Top: Evolution of the \coline\ line flux in \sneia\ with nebular time series ($\geq3$ observations past 150 days), compared to curves following the mass of \cofs\ as a function of time to the first power (dotted line) and second power (solid line).  Data for each SN was shifted by a multiplicative offset (i.e. log additive offset) that reduced residuals with the $M_{Co}(t)^2$ line.  Bottom: Evolution of the ``Fe complex'' flux with phase, compared to the same lines as above as well as an additional line proportional to the product of the \cofs\ mass with the \fefs\ mass as a function of time (dashed curve).}
\label{fig:neb_time_series}
\end{center}
\end{figure}

In the upper panel of Figure~\ref{fig:neb_time_series} we show the evolution of the \coline\ line luminosity versus time for our sample of \sneia\ with three or more observations after +150~days.  We plot the line evolution for a linear (dotted) line and quadratic (solid line) dependence on $M_{Co}(t)$, with both curves normalized at phase $t=+200$~days.  For each SN in this subset, we fit for a single multiplicative scaling factor that minimizes the residuals of the $M_{Co}(t)^2$ line (i.e. we normalize each SN data set to that line -- thus the reason for requiring multiple data points per SN).  This isolates the time dependence of the line flux (which depends on the SN nebula physics) by removing its absolute magnitude (which depends on the quantity of \nifs\ produced).

The evolution of the \coline\ line shows a remarkable agreement with the expected trend of $M_{Co}(t)^2$, perhaps as early as phase $+150$~days.
The one possible exception to the $M_{Co}(t)^2$ trend is SN~1991T, which appears to have a shallower evolution than the other \sneia.  As we show below (Section~\ref{sec:co_evo_gamma}), this cannot arise from gamma-ray opacity.  Instead the most likely explanation is probably a higher ionization state at early epochs ($t \leq 300$~days).  Because of this, for SN~1991T {\em only} we excise epochs prior to $300$~days when calculating its \nifs\ mass in Section~\ref{sec:mni_from_lco} -- a choice which yields more favorable agreement with previous analyses from the literature.

To contrast the behavior of the \coline\ line with other regions of the nebular spectra, we also inspected the evolution of the blue ``Fe complex'' of lines.  For each spectrum we integrate the flux in the region $4100-5600$~\AA\ (adjusted for each SN according to its central nebular velocity measured from the Co line) where the emission is almost entirely dominated by Fe lines (see Figure~\ref{fig:cmfgen_models}).  Following our arguments for the expectation of the \coline\ line flux, the Fe complex flux should be proportional to the energy being deposited -- which scales as $M_{Co}(t)$ -- and the mass fraction of Fe (which should be relatively constant as $M_{Co} \ll M_{Fe}$ at this point).  Thus the Fe complex flux should scale linearly with $M_{Co}(t)$.  In the lower panel of Figure~\ref{fig:neb_time_series} we plot the evolution of the Fe flux for the sample of \sneia, and see that it follows more closely the $M_{Co}(t)$ curve than the $M_{Co}(t)^2$ curve.  However, we do note deviation from this line such that the logarithmic slope is somewhat intermediate between 1 and 2.  Additionally, earlier epochs are subject to a complicated interplay of additional energy deposition from gamma-rays (as for the \coline\ line, see Section~\ref{sec:co_evo_gamma}), decreased emission due to nonzero optical depth in this region of the spectrum, and possible emission from \CoII\ (see Figure~\ref{fig:cmfgen_models}).

We note that the above results also explain one aspect of the data presented in \citet{forster13}.  Those authors examined the late ($35 \lesssim t \lesssim 80$~days) colour evolution (i.e. Lira law) for a large sample of nearby \sneia\ and its relationship with dust absorption (as inferred from narrow sodium absorption).  The mean value of $B$-band decline rates were roughly 0.015 mag/day, while the V-band decline rates were nearly twice that (0.030 mag/day).  The $B$-band is dominated by the Fe complex whose flux decays as $M_{Co}(t)$, while the V-band is heavily influenced by Co lines (see Figure~\ref{fig:cmfgen_models} in Section~\ref{sec:line_theory}) whose flux decays as $M_{Co}(t)^2$.  This naturally explains why the luminosity decay rate (in mag/day) in V-band is nearly twice that of the $B$-band, and contributes to why \sneia\ become bluer (in $B-V$) with time at these epochs.

\subsection{Testing gamma-ray opacity effects on \coline\ evolution}
\label{sec:co_evo_gamma}
While the data appear to agree with an $M_{Co}(t)^2$ dependence of the \coline\ flux evolution, it is important to investigate the impact of gamma-ray energy deposition on deviation from this parametrization.

\begin{figure}
\begin{center}
\includegraphics[width=0.48\textwidth]{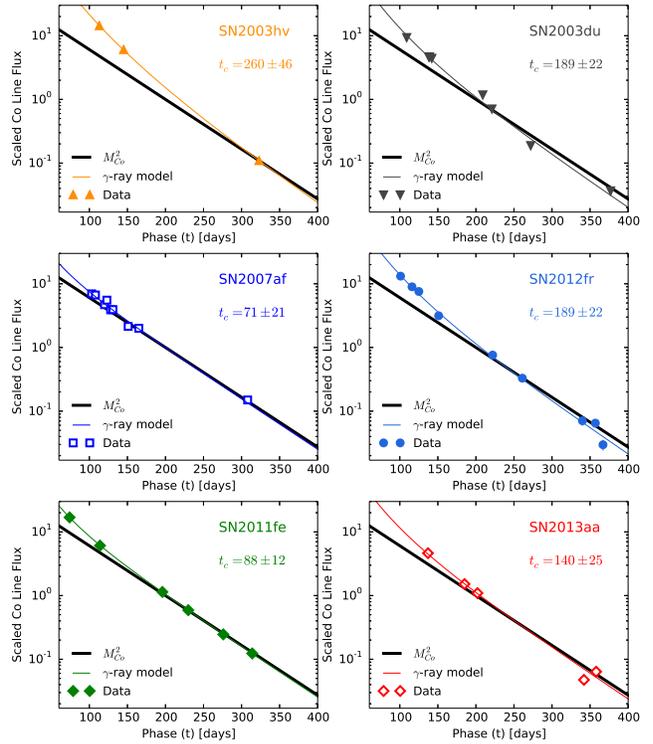}
\caption{Fits of gamma-ray opacity model to select \snia\ \coline\ line fluxes. The fitted ``crossing'' time (when gamma-ray and positron energy deposition are equal) is shown in each panel.}
\label{fig:gamma_ray_effects}
\end{center}
\end{figure}

To this end, we isolated the subset of \sneia\ from our sample with at least one nebular spectrum {\em earlier} than +150 days and at least one spectrum {\em later} than +250 days.  For the six \sneia\ satisfying these criteria, we fit the \coline\ flux evolution using the parametrization of Equation~\ref{eq:final_co_scaling}.  This fit has two free parameters: a multiplicative scaling for all the line fluxes, and the gamma-ray ``crossing'' time $t_c$ when energy deposition from gamma-rays and positrons are equal.  These fits are shown in Figure~\ref{fig:gamma_ray_effects}.

In general the \coline\ evolution is extremely well fit by this model, especially for \sneia\ with good temporal coverage and high signal-to-noise data (notably SN~2011fe and SN~2012fr).  Some \sneia\ have a gamma-ray crossing time similar to the prediction from our model ($t_c\sim200$~days) while some other \sneia\ have shorter crossing times ($t_c\sim80$~days).  The implications of this for \snia\ progenitors will be discussed in further detail in Section~\ref{sec:cofs_decay}.  Given these gamma-ray opacity model fit results, we calculate that deviations of \coline\ flux evolution from the simple $M_{Co}(t)^2$ could range from 15\% to 100\% at $t=150$, and 7\% to 55\% at $t=200$, and 4\% to 30\% at $t=250$ days.


\section{Measuring $^{56}$Ni mass from \snia\ nebular spectra}
\label{sec:nickel_mass}

\subsection{Placing \coline\ flux measurements at disparate epochs on a common scale}
\label{sec:co200_scaling}
To place all our \snia\ \coline\ fluxes on a common scale, we first convert the observed line flux to the absolute line luminosity emitted by the SN using the distance to the SN host galaxy.  For some \sneia\ in our sample, redshift-independent distance measurements exist for the host galaxy, particularly a number with Cepheid distance measurements.  For most of the \sneia\ in our sample, however, the SN distance is computed by converting the host galaxy redshift to a distance using a Hubble constant value of $H_0 = 73.8$~km~s$^{-1}$~Mpc$^{-1}$ chosen from \citet{riess11} to maintain consistency with those hosts with Cepheid distances from that work.  For hosts with redshift-based distances, we assign a distance uncertainty corresponding to a peculiar velocity uncertainty of 300~km~s$^{-1}$.  Table~\ref{tab:host_info} lists the full set of distance moduli (and references) employed in our sample.

Calculating the absolute \coline\ flux emitted by each SN also requires correction for extinction by interstellar dust in the SN host galaxy.  We accomplish this by calculating the \citet[][hereafter CCM]{ccm} reddening curve at the rest central wavelength of the \coline\ complex for an appropriate value of the reddening $E(B-V)$ and selective extinction $R_V$.  For most \sneia\ in our sample, the reddening is extremely low ($E(B-V) \leq 0.10$~mag), so we use the light curve colour fitted by SiFTO \citep{sifto}, and a selective extinction value of $R_V=2.8$ \citep[appropriate for cosmological \sneia, see][]{chotard11}.  We note that the choice of $R_V$ has negligible impact on the majority of our sample.  \snia\ light curve colours are affected by both intrinsic colour and host galaxy extinction \citep[see, e.g.,][]{scolnic14a}, so for \sneia\ with negative SiFTO colours -- indicating blue intrinsic colours -- we apply no colour correction (i.e. colour corrections never {\em redden} the data).  In this work, we are not trying to standardize SN Ia (in which applying a colour correction to the intrinsic colours may also be appropriate); rather we are only concerned with eliminating the effects of dust extinction.

Two \sneia\ in our sample, however, have strong extinction by unusual dust and thus must be treated differently.  SN~2014J occurred behind a thick dust lane in the nearby starburst galaxy M82.  \citet{foley14} performed a detailed fit of multi-colour photometry of the SN, and find it is best fit by a CCM-like reddening curve with $E(B-V)=1.19$ and $R_V=1.64$.  We adopt their colour correction for SN~2014J, and for the line flux uncertainty arising from the reddening correction we adopt their uncertainty for the visual extinction of $\sigma_{A_V} = 0.18$~mag.  SN~2007le showed moderately low extinction but with some variability in the sodium absorption feature likely arising from interaction of the SN with its circumstellar medium \citep{simon09}.  Despite this variability, {\em most} of the absorption strength remains stable, so we adopt a colour correction for SN~2007le with $E(B-V)=0.277$ and $R_V=2.56$ as derived by \citet{simon09}.

\begin{figure}
\begin{center}
\includegraphics[width=0.45\textwidth]{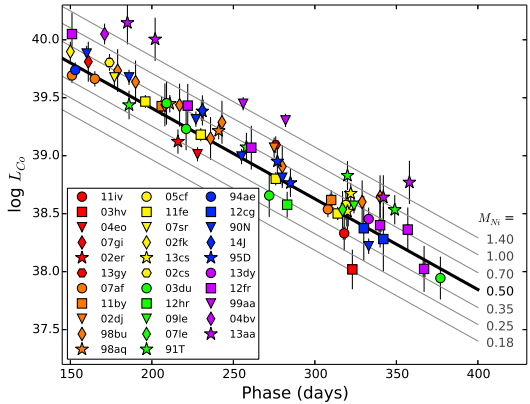}
\caption{Evolution of the {\em absolute} \coline\ line luminosity as a function of phase for all \sneia\ in our sample.  The solid line corresponds to the square of the mass of \cofs\ as a function of time, anchored by the \coline\ luminosity for SN~2011fe at $\sim200$~days. Here thick error bars correspond to flux measurement errors, while narrow error bars correspond to distance uncertainties.}
\label{fig:co_vs_phase}
\end{center}
\end{figure}

Figure~\ref{fig:co_vs_phase} presents the total emitted \coline\ luminosity as a function of phase for all nebular spectra in our final sample.  In this and subsequent figures, the thick errorbars represent the composite measurement errors from the \coline\ flux, $B$-band flux in the spectrum, observed (photometric) $B$-band magnitude, and extinction correction; the narrow error bars represent the distance uncertainties.  Points are colour-coded (in groups) based on the light curve stretch.

The line luminosity values are then used to compute an {\em effective} luminosity of the \coline\ line at a common phase of +200~days for all \sneia\ in the sample (hencefoward we refer to this as $L_{Co}$) using the $M_{Co}^2$ curve.  For a single nebular spectrum, this can be calculated directly as:
\begin{equation}
  \log(L_{Co}(200)) = \log(L_{Co}(t)) + 7.80\times10^{-3}*(t-200)
\end{equation}
For \sneia\ with multiple spectra, $L_{Co}$ is calculated as the \chisq-weighted mean value across all acceptable epochs ($150 \leq t \leq 400$~days) using the above equation.  We note the above equation is calculated assuming a time between explosion and $B$-band peak (i.e. rise time) of 17~days, but there may be an associated uncertainty on this due to diversity in \snia\ rise times \citep{ganesh10} and possible dark phase before first light escapes \citep{pn13}.  Each day of difference in explosion date results in a corresponding change in the final \coline\ luminosity of 1.8\% -- assuming an explosion date uncertainty of about 3 days, we thus expect the explosion date uncertainty contributes about 5\% uncertainty to the final nickel mass derived in Section~\ref{sec:mni_from_lco}.

As noted in Section~\ref{sec:neb_time_series}, SN~1991T may represent a case where the stable ionization state is not established until later than other SNe (also evident in Figure~\ref{fig:co_vs_phase}), so for this SN we use the later two epochs ($t \geq 300$) to establish  $L_{Co}$.  This also yields a favorable agreement of our \nifs\ mass with literature estimates (see Section~\ref{sec:mni_from_lco}).

\begin{figure}
\begin{center}
\includegraphics[width=0.45\textwidth]{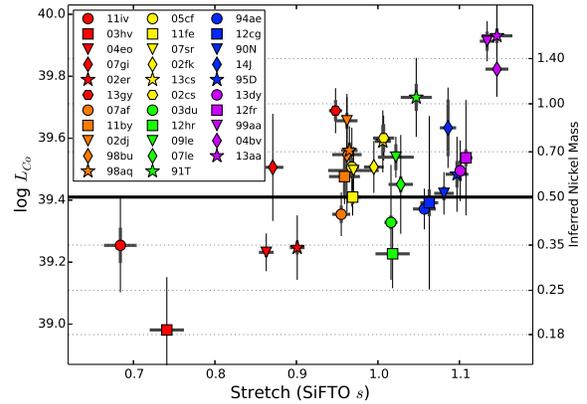}
\caption{\coline\ line luminosity scaled to its equivalent value at $t=200$~days using the $M_{Co}(t)^2$ curve ($L_{Co}$) versus SN light curve stretch.  As in Figure~\ref{fig:co_vs_phase}, thick error bars correspond to flux measurement errors, while narrow error bars corresponding to distance uncertainty arising from peculiar velocities.}
\label{fig:co_vs_stretch}
\end{center}
\end{figure}

In Figure~\ref{fig:co_vs_stretch} we show the scaled $t=200$~d \coline\ line luminosity plotted against light curve stretch.  A clear correlation is evident between the \coline\ line luminosity and stretch -- this is expected given the \coline\ luminosity traces the amount of \nifs\ produced in the explosion, and \nifs\ directly powers the peak luminosity which correlates with the light curve stretch.

\subsection{Inferring \mni\ from \coline\ flux}
\label{sec:mni_from_lco}
Scaling the \coline\ flux values to the same phase ($t=200$~days) effectively places all measurements at the same epoch since explosion, so the amount of \cofs\ will have the same proportionality to the amount of \nifs\ produced in the explosion.  The final critical ingredient for inferring \cofs\ mass (and thus \nifs\ mass) from the \coline\ line flux is the scaling between \cofs\ mass and \coline\ flux.  For reasons we will explore in Section~\ref{sec:nebular_stability}, we expect this conversion factor to be relatively stable in time (for phases $150 \leq t \leq 400$~days considered here) and consistent across all \sneia.  At these phases we also expect \cofs\ to be the dominant isotope (by mass) of cobalt \citep{seitenzahl09}, as $^{57}$Co only dominates energy deposition around $t\sim1000$~days \citep{graur15}.

We expect the \coline\ line flux at phase $t=200$~days to be linearly proportional to the mass of \nifs\ produced in explosion (since the \cofs\ mass fraction at this same epoch is necessarily the same for all \sneia).  To convert \coline\ flux to \nifs\ mass requires some scaling between the two quantities to be determined.  In principle this could be computed through radiative transfer modelling of late phases for \snia\ explosion models.  However, for simplicity in this work, we choose to anchor the relation with the well-studied \snia\ SN~2011fe.  Modelling of the photospheric phase light curve for SN~2011fe by \citet{pereira13} yielded a \nifs\ mass of $M_{Ni} = 0.53 \pm 0.11 M_\odot$.  Recently \citet{mazzali15} extended their spectroscopic modelling of the SN~2011fe spectral time series \citep[presented for photospheric epochs in][]{mazzali14} to nebular phase epochs and found $M_{Ni} = 0.47 \pm 0.08 M_\odot$.  For simplicity in this work, we thus will choose a \nifs\ mass anchor for SN~2011fe of $M_{Ni} = 0.50 M_\odot$, yielding final \nifs\ mass values derived as:
\begin{equation}
  M_{Ni} = 0.50~M_\odot\frac{L_{Co}}{L_{11fe}}
\end{equation}
where $\log(L_{11fe}) = 39.410$ is the scaled \coline\ luminosity we measure for SN~2011fe -- this is used as a zeropoint for the remainder of our SN sample.  The values for \mni\ for our sample are presented in Table~\ref{tab:nickel_masses}.  In Section~\ref{sec:mni_vs_mej} we further discuss the implications of our \nifs\ mass values and their relation to the ejected masses of our \snia\ sample.

\begin{table}
\begin{center}
\caption{Final SN Ia Nickel Masses}
\label{tab:nickel_masses}
\begin{tabular}{lll}
\hline
SN     & $M_{Ni}$         & $M_{ej}$         \\
       & ($M_\odot$) $^a$ & ($M_\odot$) $^b$ \\
\hline
SN1990N   & $0.514 \pm 0.027(0.081)$ & $1.437 \pm 0.009$ \\
SN1991T   & $1.049 \pm 0.106(0.308)$ & $1.407 \pm 0.019$ \\
SN1994ae  & $0.458 \pm 0.013(0.069)$ & $1.417 \pm 0.013$ \\
SN1995D   & $0.593 \pm 0.059(0.165)$ & $1.448 \pm 0.009$ \\
SN1998aq  & $0.707 \pm 0.042(0.127)$ & $1.304 \pm 0.015$ \\
SN1998bu  & $0.686 \pm 0.029(0.292)$ & $1.299 \pm 0.027$ \\
SN1999aa  & $1.593 \pm 0.114(0.238)$ & $1.465 \pm 0.003$ \\
SN2002cs  & $0.775 \pm 0.081(0.130)$ & $1.361 \pm 0.016$ \\
SN2002dj  & $0.882 \pm 0.051(0.176)$ & $1.299 \pm 0.019$ \\
SN2002er  & $0.344 \pm 0.018(0.082)$ & $1.202 \pm 0.015$ \\
SN2002fk  & $0.625 \pm 0.016(0.120)$ & $1.346 \pm 0.016$ \\
SN2003du  & $0.414 \pm 0.022(0.177)$ & $1.373 \pm 0.010$ \\
SN2003hv  & $0.186 \pm 0.003(0.073)$ & $0.914 \pm 0.037$ \\
SN2004bv  & $1.294 \pm 0.040(0.266)$ & $1.468 \pm 0.003$ \\
SN2004eo  & $0.332 \pm 0.011(0.046)$ & $1.135 \pm 0.016$ \\
SN2005cf  & $0.625 \pm 0.044(0.184)$ & $1.308 \pm 0.013$ \\
SN2007af  & $0.440 \pm 0.029(0.071)$ & $1.289 \pm 0.017$ \\
SN2007gi  & $0.624 \pm 0.027(0.248)$ & $1.149 \pm 0.023$ \\
SN2007le  & $0.549 \pm 0.033(0.202)$ & $1.387 \pm 0.017$ \\
SN2007sr  & $0.609 \pm 0.027(0.107)$ & $1.311 \pm 0.045$ \\
SN2009le  & $0.673 \pm 0.065(0.102)$ & $1.380 \pm 0.026$ \\
SN2011by  & $0.582 \pm 0.082(0.119)$ & $1.295 \pm 0.029$ \\
SN2011fe  & $0.500 \pm 0.026(0.069)$ & $1.310 \pm 0.015$ \\
SN2011iv  & $0.349 \pm 0.046(0.122)$ & $0.818 \pm 0.032$ \\
SN2012cg  & $0.479 \pm 0.048(0.309)$ & $1.422 \pm 0.010$ \\
SN2012fr  & $0.670 \pm 0.043(0.287)$ & $1.454 \pm 0.004$ \\
SN2012hr  & $0.328 \pm 0.008(0.084)$ & $1.375 \pm 0.025$ \\
SN2013aa  & $1.658 \pm 0.091(0.717)$ & $1.468 \pm 0.004$ \\
SN2013cs  & $0.757 \pm 0.094(0.174)$ & $1.360 \pm 0.013$ \\
SN2013dy  & $0.608 \pm 0.047(0.137)$ & $1.450 \pm 0.004$ \\
SN2013gy  & $0.950 \pm 0.075(0.159)$ & $1.278 \pm 0.012$ \\
SN2014J   & $0.837 \pm 0.176(0.250)$ & $1.441 \pm 0.007$ \\
\hline
\end{tabular}
\end{center}
$^a$ Nominal uncertainties arise from measurement errors in the Co line flux or SN reddening, while distance uncertainties are listed in parenthesis.  Systematic error for $M_{Ni}$ is estimated at $0.2M_\odot$.\\
$^b$ Includes only measurement uncertaintes from SN light curve stretch.  Systematic error for $M_{ej}$ is estimated at $0.1M_\odot$. \\
\end{table}

\begin{figure}
\begin{center}
\includegraphics[width=0.45\textwidth]{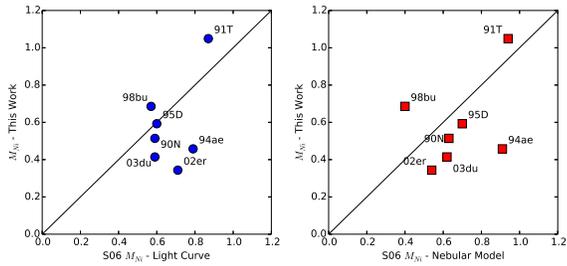}
\caption{Comparison of our \nifs\ mass values to those derived by \citet{stritzinger06b} using light curve modelling (left) and nebular spectra modelling (right).}
\label{fig:mni_checks}
\end{center}
\end{figure}

Other techniques have been presented for measuring the mass of \nifs\ produced in the \snia\ explosion.  \citet{stritzinger06} employed semi-empirical modelling of \snia\ bolometric light curves to measure the ejected mass and \nifs\ mass for a sample of 17 nearby \sneia.  They then found that \nifs\ masses derived from modelling of the nebular spectra \citep{mazzali97, mazzali98, stehle05} yielded consistent results \citep{stritzinger06b}.  Seven of the \sneia\ from their sample are included in ours, and we show a comparison of our \nifs\ values versus those derived from their two methods in Figure~\ref{fig:mni_checks}.  In some of the cases, our \nifs\ masses are somewhat lower than theirs (both for the light curve and nebular \nifs\ mass estimates) though generally show acceptable agreement.  We note that for SN~1994ae and SN~2002er , \citet{stritzinger06b} employ a much higher reddening value than ours ($E(B-V)=0.15$~mag versus $E(B-V)=0.00$~mag for SN~1994ae and $E(B-V)=0.36$~mag versus $E(B-V)=0.12$~mag for SN~2002er), which is likely the source of the discrepancy between our values.

\section{Discussion}
\label{sec:discussion}
In this Section we discuss the important physical implications of our observational results above.  First, we examine the fact that the \coline\ line flux evolution requires a constant scaling between energy released by \cofs\ decay and that absorbed by the nebula -- this requires efficient local deposition of energy from positrons and near-complete escape of gamma-rays from the IGE core (Section~\ref{sec:cofs_decay}).  Next, we argue that the \coline\ evolution requires stable ionization conditions in the nebula for a period of several hundred days, which we support by demonstrating stability of ionization-dependent flux ratios measured from the data (Section~\ref{sec:nebular_stability}).  Finally, we discuss potential interpretations of \snia\ explosion conditions implied by our observed relationship between inferred \nifs\ mass and ejected mass (Section~\ref{sec:mni_vs_mej}).

\subsection{Gamma-ray transparency timescales for nebular \sneia}
\label{sec:cofs_decay}
For \cofs\ to deposit a constant fraction of its decay energy into the nebula, positrons from the decay must be efficiently trapped in the IGE core and gamma-rays must be able to effectively escape\footnote{We do note that other physical properties of the nebula (e.g. ionization or emission measure changes) could somehow conspire to compensate for gamma-ray opacity to make the line emission evolve as $M_{Co}^2$, but we consider the gamma-ray transparency scenario to be the simplest explanation.}.  As noted above, efficient local positron energy deposition is expected to hold for the temperatures and densities encountered at these nebular phases \citep{axelrod80, chan93, rs98}.  In practice, gamma-rays become negligible after the time when the gamma-ray energy deposition equals that from positrons, which occurs when the optical depth drops enough to reach this equality (Section~\ref{sec:co_evo_theory}).  We will refer to this henceforth as the ``transparency'' timescale $t_c$ -- after this epoch positrons dominate energy deposition in the nebula.

We fit the transparency timescale for several supernovae in Section~\ref{sec:co_evo_gamma} and found several have longer transparency times ($t_c\sim180$~days) close to the theoretical expectation for the DDC10 model.  Previous analysis of gamma-ray transparency timescales found similar results: $t_c\approx170$~days for SN~2000cx \citep{sollerman04} and $t_c\approx188$~days for SN~2001el \citep{ss07}.  However we found that other \sneia\ (notably SN~2011fe and SN~2007af) had much shorter transparency times ($t_c\sim80$~days).  This variation in transparency times may reflect a diversity in nebular densities, as most of the gamma-ray opacity at these late epochs will come from opacity from electrons in the nebula.  Interestingly, the \sneia\ with shorter transparency times (SN~2011fe and SN~2007af) have lower stretch values than most of the \sneia\ with longer transparency times (SN~2003du, SN~2012fr, SN~2013aa), possibly indicating some relationship between nebular density and stretch.  The one exception to this is SN~2003hv, which appears to have low stretch but long transparency time (and thus would imply high density) -- this result is opposite to the findings of \citet{mazzali11} who found SN~2003hv had reduced density in the inner regions of the ejecta.  The source of this discrepancy is unclear, but may constitute further evidence that SN~2003hv is a ``non-standard'' event.

Because of the diversity in gamma-ray transparency timescales in the \sneia\ we tested, it is likely that the impact of gamma-ray energy deposition on the \coline\ flux will be impacted by similar variability.  Given the results above (Section~\ref{sec:co_evo_gamma}) this may result in an average uncertainty of 30\% on the final \nifs\ masses we infer.  The only robust way to account for gamma-ray opacity effects is to obtain a nebular time series.  However the transparency time is best constrained by observations from 100-150 days when the SN is only 3-4 magnitudes fainter than peak.  Thus it should be observationally feasible to obtain such data for future \sneia\ observed in the nebular phase.

More interestingly, the time evolution of the \coline\ flux presents a new method for measuring the gamma-ray transparency time scale, as it gives a direct probe of the energy being deposited into the nebula.  Previously this could only be done with the aid of bolometric light curves \citep{sollerman04, ss07, leloudas09}, which necessarily rely on extensive optical and infrared photometry and/or uncertain bolometric corrections.  Instead, our method requires only two nebular spectra with contemporaneous optical photometry.

\subsection{Ionization conditions in the SN nebula}
\label{sec:nebular_stability}
As noted above, the consistency of the \coline\ flux evolution with the square of the cobalt mass implies a constant scaling between the energy being absorbed by cobalt atoms and the energy they emit in the \coline\ line.  This implies stability in the ionization conditions of the nebula, which we now investigate from a more detailed inspection of our nebular spectra.

\begin{figure*}
\begin{center}
\includegraphics[width=0.95\textwidth]{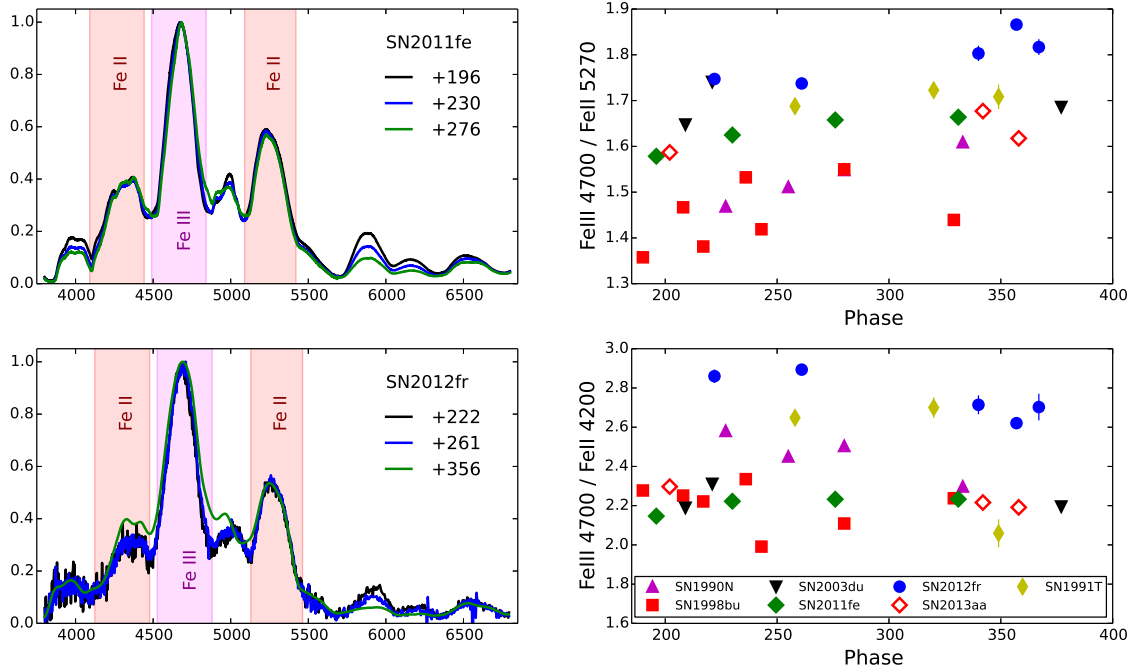}
\caption{Left panels: Multiple nebular phase spectra of SN~2011fe (top) and SN~2012fr (bottom), highlighting the flux integration regions for the line complexes dominated by \FeII\ (light red regions) and \FeIII\ (light magenta regions).  Integration zones are shifted by the central redshift of the nebular Fe lines, here a blueshift of $\sim600$~\kms\ for SN~2011fe and a redshift of $\sim1800$~\kms\ for SN~2012fr.  Right panels: Temporal evolution of the ratios of the flux integral for the \FeIII\ 4700~\AA\ complex compared to the \FeII\ 5270~\AA\ complex (top) and \FeII\ 4200~\AA\ complex (bottom) for \sneia\ with nebular time series.}
\label{fig:fe_ratio_vs_phase}
\end{center}
\end{figure*}

To confirm that the ionization state of the nebula is indeed slowly evolving from phases $150 \leq t \leq 400$ days, we examine the flux ratios of nebular emission lines arising primarily from \FeII\ and \FeIII.  If the ratio of these lines evolves with time, this would indicate a change in the ionization state.  In the left panels of Figure~\ref{fig:fe_ratio_vs_phase} we highlight the regions of the typical \snia\ nebular spectra (here from SN~2011fe and SN~2012fr) which are dominated by strong line complexes of either \FeII\ or \FeIII.  We integrate the flux in these regions for all the nebular \snia\ spectra in our sample, and in the right panels of Figure~\ref{fig:fe_ratio_vs_phase} we show how the line flux ratios evolve with phase for the nebular time series \sneia\ (the same as from Section~\ref{sec:neb_time_series}).  For this analysis we only consider phases later than $t\sim200$~days, as this is when this region of the spectrum is reliably optically thin (see Section~\ref{sec:neb_time_series}) -- note this cuts SN~2007af from the Fe time series sample.

Though there is indeed some evolution in the flux ratio of \FeII\ lines to \FeIII\ lines, it is comparatively small -- generally less than 10\% change of the relative line flux in \FeIII\ compared to \FeII.  In sharp contrast, consider the \FeIII/\FeII\ line flux ratios as measured from the $t\sim1000$~days spectrum for SN~2011fe from \citet{tauben15} -- 0.52 for 4700/5270 versus a mean of 1.6 at earlier phases, and 0.87 for 4700/4200 versus an earlier mean of 2.3 -- which decrease by at least 65\% from their values in the $150 \leq t \leq 400$ day range [we note these values should be considered upper limits as it appears that the \FeIII\ 4700 line has effectively disappeared in the $t\sim1000$~days spectrum for SN~2011fe, so the flux we measure here is likely due to other species].  By these very late phases the physical conditions in the \snia\ nebula have clearly changed in a dramatic fashion.  Such is not the case for the \sneia\ in our sample at phases $150 \leq t \leq 400$ days.

\begin{figure}
\begin{center}
\includegraphics[width=0.45\textwidth]{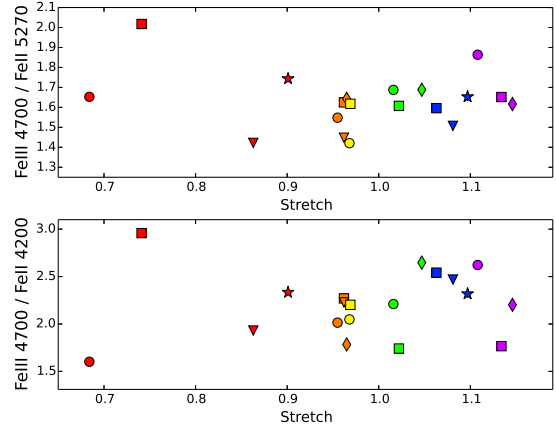}
\caption{Integrated flux ratios of the \FeIII\ 4700~\AA\ complex compared to the \FeII\ 5270~\AA\ complex (top) and \FeII\ 4200~\AA\ complex (bottom) as a function of light curve stretch (SiFTO $s$) for all \sneia\ in our sample.  Formal spectrum flux error bars are smaller than the data markers.  Markers are the same as for Figure~\ref{fig:co_vs_stretch}}
\label{fig:fe_ratio_vs_stretch}
\end{center}
\end{figure}

In order to meaningfully compare the \coline\ line flux from different \sneia, another key requirement is that the ionization state of {\em all} \sneia\ be relatively similar.  To test this assumption, we again use the Fe line ratios described above, but plot the mean \FeIII/\FeII\ line flux ratio (computed as the error-weighted mean for \sneia\ with multiple epochs) versus light curve stretch in Figure~\ref{fig:fe_ratio_vs_stretch}.  We have excluded the highly reddened SN~2007le and SN~2014J to avoid any biases in these ratios due to uncertainty in the dust law (i.e. $R_V$).

Here we see some mild coherent change in the Fe line flux ratios (and thus ionization state) as a function of light curve stretch \citep[with SN~2003hv as an outlier, as previously noted by][]{mazzali11}.  Here the overall range of the line ratios is somewhat larger, with variations perhaps up to 40\% but with a scatter of 7\% (for 4700/5270) and 15\% (for 4700/4200).  The ionization potentials of Fe and Co are very similar, which means a change in \FeIII\ line flux induced by variation of the ionization state will manifest a comparable change in \CoIII\ line flux.  Thus our \coline\ line fluxes above should have an additional scatter due to ionization state variations of about 10\%.  Since our inferred \nifs\ masses are proportional to this line flux, this means that ionization state variations could induce a scatter of similar magnitude in our \nifs\ masses.

Our measurement of the \coline\ line flux evolution, and variations of \FeIII/\FeII\ line flux ratios as a function of both phase and SN stretch, coherently indicate that the ionization states of normal \sneia\ are remarkably consistent across different SNe and nearly constant across phases $150 \leq t \leq 400$ days.  This stability of the ionization state was predicted by \citet{axelrod80}, and our results here present the most compelling evidence to-date in support of that prediction.

\subsection{The relationship between \nifs\ and Ejected Mass}
\label{sec:mni_vs_mej}
The relationship between \coline\ luminosity and light curve stretch (Figure~\ref{fig:co_vs_stretch}) hints at a relationship between physical properties of the \snia\ progenitor system.  In Section~\ref{sec:mni_from_lco} we converted our measured \coline\ line luminosities into inferred \nifs\ masses.  Here we convert light curve stretch into the SN ejected mass (i.e. progenitor mass for \sneia) using the relationship between \mej\ and light curve stretch discovered by \citet{scalzo14a}.  \citet{scalzo14c} used Bayesian inference to model the intrinsic distribution of ejected masses, which can be folded in as an additional prior when determining ejected mass using this relation.  We derive a cubic fit to the relationship between stretch and \mej:
\begin{eqnarray}
  M_{ej} &=& 2.07 - 7.51s + 11.56s^2-4.77s^3 \\
        &=& 1.35 + 1.30 (s-1) - 2.75 (s-1)^2 - 4.77 (s-1)^3 \nonumber
\end{eqnarray}
The resultant values for ejected mass (\mej) we derive are presented in Table~\ref{tab:nickel_masses} along with our \nifs\ masses.

\begin{figure*}
\begin{center}
\includegraphics[width=0.95\textwidth]{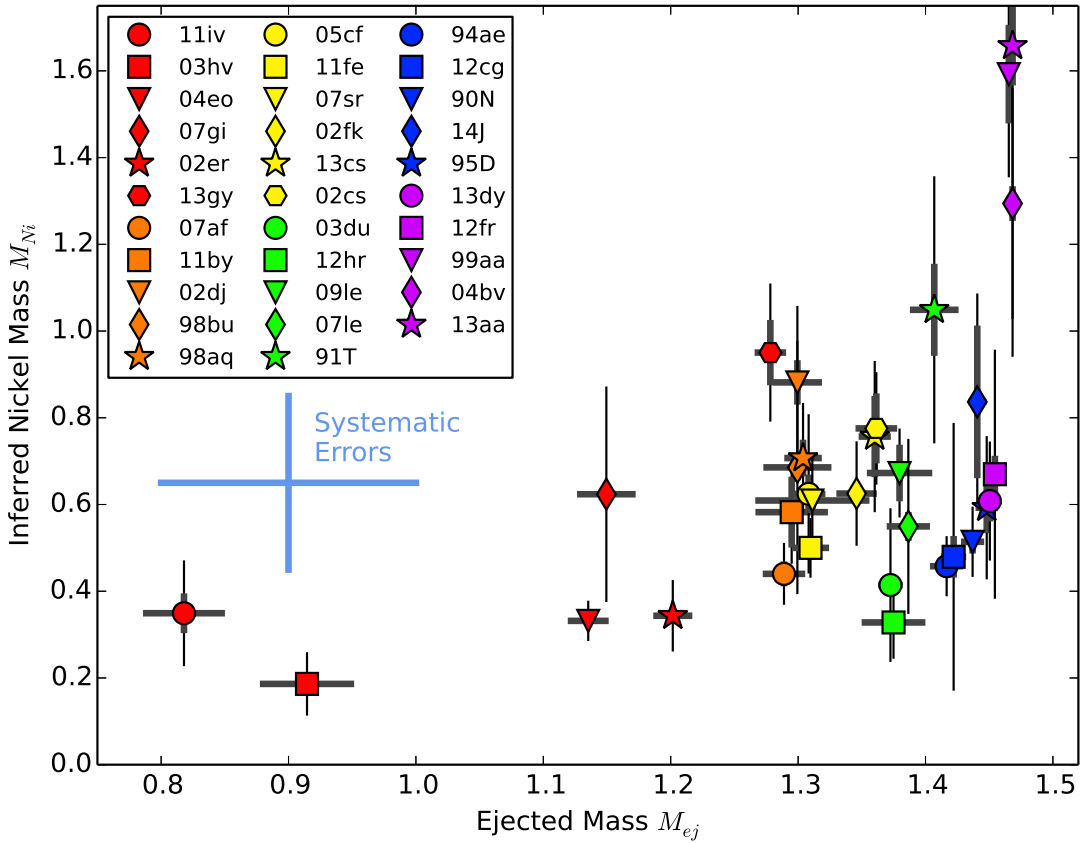}
\caption{Mass of \nifs\ inferred from the scaled \coline\ line luminosity versus total ejected mass (i.e., progenitor WD mass) inferred from SN light curve stretch. As in Figure~\ref{fig:co_vs_phase}, thick error bars correspond to flux measurement errors (including colour correction uncertainties), while narrow error bars corresponding to distance uncertainties arising from peculiar velocities. The typical systematic uncertainties for estimating \mej\ \citep[$\pm0.1M_\odot$ - from the stretch-\mej\ relation of][]{scalzo14c} and \mni\ ($\pm0.2M_\odot$ - from gamma-rays, explosion date uncertainty, and possible line contamination) are shown as the blue error bars in the left side of the plot.  Note the anomalously high \mni\ values for SN~1999aa and SN~2013aa, which we attribute to line contamination and distance uncertainty, respectively (see text for details).}
\label{fig:mni_vs_mej}
\end{center}
\end{figure*}

In Figure~\ref{fig:mni_vs_mej} we plot our inferred \nifs\ masses against these ejected masses. We note that there is a systematic uncertainty associated with \mej\ calculation of about $0.1M_\odot$, as determined by \citet{scalzo14a} from recovering masses of \snia\ explosion models.  For \nifs\ masses, we previously noted several sources of uncertainty: 10\% uncertainty in the \coline\ flux itself (Section~\ref{sec:line_measure}), 5\% uncertainty on the $t=200$ \coline\ luminosity due to uncertainty in the explosion date (Section~\ref{sec:co200_scaling}), 10\% from ionization state variations (Section~\ref{sec:nebular_stability}), and possibly 30\% from variations in gamma-ray transparency timescales (Section~\ref{sec:cofs_decay}).  Collectively this constitutes a possible 35\% uncertainty in our \nifs\ masses, which given the values we find would produce a mean uncertinty in \mni\ of about $0.2M_\odot$.

The relation between \mni\ and \mej\ shows potential evidence for two regimes for the production of \nifs\ in \sneia.  For sub-Chandrasekhar ejected masses ($M_{ej} \lesssim 1.3M_\odot$ -- though note SN~1991bg-like objects are not included in this analysis), the amount of \nifs\ produced is clustered around $M_{Ni} \sim 0.4M_\odot$, with a possible increase of \mni\ with \mej\ (though we note the statistics are small).  Chandrasekhar-mass progenitors ($M_{ej} \approx 1.4 \pm 0.1 M_\odot$) produce \nifs\ masses ranging from $0.4M_\odot \lesssim M_{Ni} \lesssim 1.2M_\odot$, with the extreme high \nifs\ masses  ($M_{Ni} \gtrsim 1.0M_\odot$) occuring in \sneia\ spectroscopically similar to the peculiar SNe SN~1991T (SN~1999aa, SN~2004bv, SN~2013aa, and SN~1991T itself).
Recently, \citet{fisher15} suggested that Chandrasekhar-mass \snia\ progenitors preferentially lack a vigorous deflagration phase following the initial ignition, and result in a nearly pure detonation that produces about $1.0M_\odot$ of \nifs\ and shows similarity to SN~1991T.  Our findings that the \coline\ luminosity is exceptionally high only in 91T-like \sneia\ could lend support to this theory.

We note that SN~1999aa and SN~2013aa have anomalously high \mni\ values (indeed exceeding their \mej\ values).  We visually inspected the spectra of these SNe, and find no fault in our fits to the \coline\ line.  SN~1999aa notably has the broadest linewidth of our sample, which could result in contamination of our measured \coline\ flux by nearby \FeII\ lines (see Figure~\ref{fig:cmfgen_models}).  SN~2013aa has a relatively uncertain distance to its host galaxy.  We expect the true \mni\ for these two SNe is likely to be closer to that of the other SN~1991T-like \sneia, near $1.0--1.2M_\odot$.

\begin{figure}
\begin{center}
\includegraphics[width=0.47\textwidth]{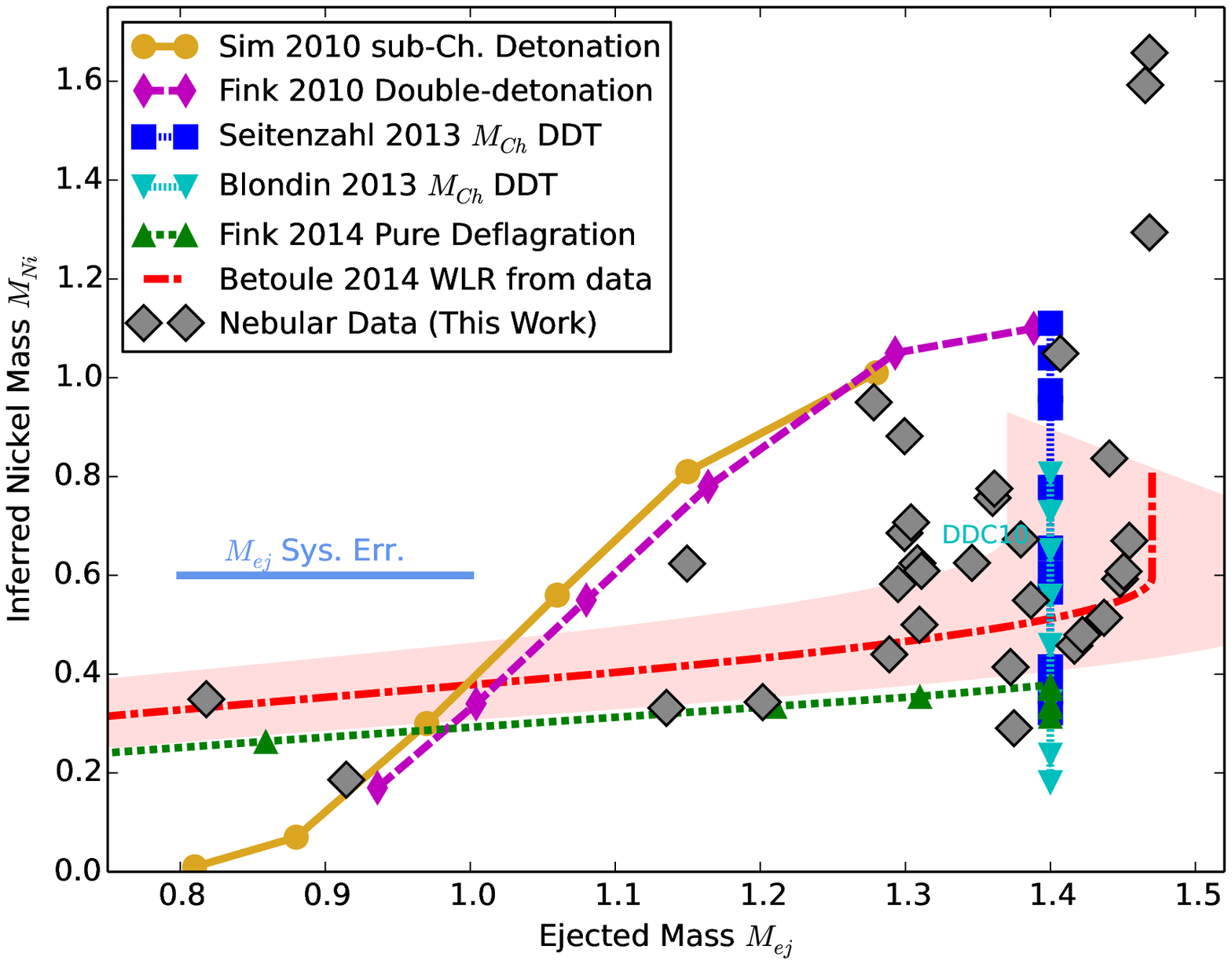}
\caption{\mni\ and \mej\ values inferred from data (dark grey diamonds -- data as in Figure~\ref{fig:mni_vs_mej}) compared to various theoretical models: pure detonation models from sub-Chandra detonations \citep[][yellow circles]{sim10} and double-detonations \citep[][magenta diamonds]{fink10}; detonation-deflagration transitions (DDT) from \citet[][blue sqaures]{seitenzahl13} and \citet[][cyan downward triangles]{blondin13}; and Chadra-mass pure deflagrations \citep[][green upward triangls]{fink14}.  The observed width-luminosity relation (and its scatter) from the recent cosmology analysis of \citet{betoule14} are shown as the red curve (and light red shaded area).}
\label{fig:models_mni_mej}
\end{center}
\end{figure}

To compare model predictions with our inferred \nifs\ mass values, we gather ejected mass and \nifs\ mass outcomes from numerous \snia\ explosion models and plot them against our data in Figure~\ref{fig:models_mni_mej}.  These models can be generally grouped into three categories: sub-Chandrasekhar mass detonations, Chandrasekhar-mass deflagration to detonation transitions (DDT), and Chandrasekhar-mass deflagrations which fail to detonate.  We discuss each category and its agreement with the data below.

{\bf Sub-Chandrasekhar (sub-Ch) mass detonations:}  We consider sub-Ch detonations from \citet{sim10}, where detonations were artificially initiated in WDs of varying initial masses.  These models are also applicable to sub-Ch WDs ignited via other mechanisms (e.g. a violent merger), and were also employed to estimate the brightness distribution of violent mergers in \citet{ruiter13}.  We also examine sub-Ch double detonation models from \citet{fink10}: these are qualitatively similar to the \citet{sim10} but the ignition mechanism naturally arises from a surface helium layer ignition.  Both models show a similar relationship between \mni\ and \mej, which shows a much steeper increase of \mni\ with \mej\ than we infer from our data.  However we note that with the systematic uncertainty in \mej\ estimates (from stretch) these may be compatible with the data.

{\bf Deflagration to detonation transitions (DDT):}  We present models from both \citet{seitenzahl13} and \citet{blondin13} -- including the DDC10 model employed for radiative transfer calculations in Section~\ref{sec:line_theory}.  In general for these models the \mch\ progenitors undergo an initial deflagration phase which transitions to a detonation at a later time: the timing of this transition directly sets the amount of \nifs\ produced.  For \citet{seitenzahl13}, the DDT time was calculated from the sub-grid scale turbulent energy \citep{ciaraldi13} which in practice varied with the vigorousness of the initial deflagration (set by hand as the number of initial ignition points).  For \citet{blondin13}, the DDT time is set by a manual trigger.  Both sets of models cover a range of \nifs\ mass production, similar to the range inferred from our data.

{\bf Pure deflagrations:}  Finally we consider pure deflagration models presented in \citet{fink14}.  These models are variations on the Chandrasekhar-mass \citet{seitenzahl13} models in which the DDT module has been intentionally turned off.  Many of these deflagration models fail to fully unbind the star and eject only a portion of the WD's total mass and leave a bound remnant -- we note that the \citet{scalzo14c} method for estimating ejected mass from light curves is not trained to account for bound remants so may have some additional systematic uncertainty for this explosion mechanism.  Interestingly, these models show a weak dependence of \mni\ on \mej\ for sub-Ch ejected masses, similar to what we infer for this regime of the data.  This also shows some agreement with the width-luminosity relation observed in cosmological supernova samples -- we show this as well in Figure~\ref{fig:models_mni_mej} using the WLR from \citet{betoule14} converted to \mni\ and \mej\ using the relations presented in \citet{scalzo14c}. 

If indeed the \mni--\mej\ trend arises from two distinct explosion mechanisms for \sneia, several key questions remain to be answered with future research.
One such question is where the split between the two mechanisms occurs -- \sneia\ at \mch\ with $\sim0.5M_\odot$ of \nifs\ could arise from either mechanism -- and what physical property of the progenitor decides which mechanism occurs.
Next we should investigate why the two mechanisms produce \sneia\ which obey the same width-luminosity relation -- one might expect that a different relationship between \mni\ and \mej\ would yield different relationship between peak luminosity and light curve width.
Such insights could be further advanced by study of other related thermonuclear explosions which span a broader range of \mni\ and \mej\ \citep[e.g.,][see their Figure 15]{mccully14a}.

Finally, we most critically should assess whether the two mechanisms calibrate cosmological distances in the same fashion.  Recent evidence has been mounting that \sneia\ show progenitor signatures (e.g. CSM interaction, high-velocity features, host galaxy properties) which appear to clump into two groups \citep{maguire14}.  In parallel, \snia\ cosmological analyses have found that \sneia\ in high- and low-mass galaxies have subtly different standardized luminosities \citep{sullivan10, kelly10, lampeitl10, gupta11, dandrea11, konishi11, galbany12, hayden13, johansson13, childress13b, rigault13, cwz14, kelly15b}.  These and the current study motivate further examination of the environments and standardized luminosities of \sneia\ whose \nifs\ mass and ejected mass are assessed with the techniques presented here.  Such a study is limited by distance uncertainties, and thus should be targeted at \sneia\ in the nearby smooth Hubble flow ($z \geq 0.015$) where distance uncertainties from peculiar velocities become small ($\leq0.10$~mag).

\section{Conclusions}
\label{sec:conclusions}
In this work we examine the \coline\ feature in \ntotspec\ nebular phase ($150 \leq t \leq 400$~days past peak brightness) spectra of \ntotsne\ \sneia\ compiled from the literature and new observations.  This feature arises predominantly from radioactive \cofs, the decay product of \nifs\ (which powers the bright early light curve) -- thus this feature provides a direct window for investigating the power source behind \snia\ light curves.

We used nebular time series for eight \sneia\ to show that the temporal evolution of the \coline\ flux falls very close to the square of the mass of \cofs\ as a function of time.  This is the expected dependence in the limit where the nebula is fully optically thin to gamma-rays produced in the \cofs\ decay but locally thermalizes energy from positrons emitted in the decay.
We then used this uniform time dependence to infer the relative amount of \nifs\ produced by all \ntotsne\ \sneia\ in our sample by using SN~2011fe as an anchor (at $M_{Ni}=0.5M_\odot$).

The greatest systematic uncertainty in our \nifs\ mass measurements was the time at which the nebula becomes effectively optically thin to gamma-rays (which we define by the ``crossing'' time when energy deposition from positrons begins to exceed that of gamma-rays).  Though this could intoduce 30\% uncertainty in \nifs\ masses (on average, though this is time dependent), we showed that the gamma-ray transparency time can be readily measured when multiple nebular spectra are available.  In particular, a single spectrum at phases $100 \leq t \leq 150$ days past maximum light -- when the SN is only 3-4 magnitudes fainter than peak -- can easily constrain the gamma-ray transparency time.  This can robustify our technique for measuring \nifs\ masses of future \sneia, but the gamma-ray transparency time itself could provide important clues to \snia\ progenitor properties.

When comparing our inferred \nifs\ masses to the ejected masses of our \snia\ sample \citep[using techniques from ][]{scalzo14a, scalzo14c}, we find evidence for two regimes in the production of \nifs\ (which are too distinct to be an artefact of systematic uncertainties).  For low ejected masses (low stretch), \mni\ clusters at low values ($M_{Ni} \approx 0.4M_\odot$).  At high ejected masses (high stretch) near the Chandrasekhar mass, \mni\ has a much larger spread ($0.4 M_\odot \lesssim M_{Ni} \lesssim 1.2 M_\odot$).  This could constitute evidence for two distinct explosion mechanisms in \sneia.

This work has illustrated the power of the nebular \coline\ feature in probing the fundamental explosion physics of \sneia.  We provide a simple recipe for calculating \nifs\ mass from the \coline\ line flux from a single nebular epoch, as well as prescription for a more robust measurement that accounts for opacity effects by using multiple nebular epochs.  Future measurements which eliminate the opacity systematics and distance uncertainties could provide a detailed understanding of the explosion mechanisms for \sneia.

\vskip11pt
{\em Acknowledgements:}
This research was conducted by the Australian Research Council Centre of Excellence for All-sky Astrophysics (CAASTRO), through project number CE110001020. BPS acknowledges support from the Australian Research Council Laureate Fellowship Grant FL0992131. ST acknowledges support by TRR33 ``The Dark Universe'' of the German Research Foundation (DFG).  DJH acknowledges support from NASA theory grant NNX14AB41G.  KM is supported by a Marie Curie Intra-European Fellowship, within the 7th European Community Framework Programme (FP7).  NER acknowledges the support from the European Union Seventh Framework Programme (FP7/2007-2013) under grant agreement n. 267251 ``Astronomy Fellowships in Italy” (AstroFIt).  MF acknowledges support from the European Union FP7 programme through ERC grant number 320360.  MS acknowledges support from the Royal Society and EU/FP7-ERC grant n$^{\rm o}$ [615929].  AGY is supported by the EU/FP7 via ERC grant no. 307260, the Quantum Universe I-Core program by the Israeli Committee for planning and budgeting and the ISF; by Minerva and ISF grants; by the Weizmann-UK ``making connections" program; and by Kimmel and ARCHES awards.  AMG acknowledges financial support by the Spanish \textit{Ministerio de Econom\'ia y Competitividad} (MINECO) grant ESP2013-41268-R.  The research leading to these results has received funding from the European Union Seventh Framework Programme [FP7/2007-2013] under grant agreement num. 264895.  SJS acknowledges funding from the European Research Council under the European Union's Seventh Framework Programme (FP7/2007-2013)/ERC Grant agreement n$^{\rm o}$ [291222] and STFC grants ST/I001123/1 and ST/L000709/1.

Part of this research was conducted while John Hillier was a Distinguished Visitor at the Research School of Astronomy and Astrophysics at the Australian National University.
This research is based on observations collected at the European Organisation for Astronomical Research in the Southern Hemisphere, Chile as part of PESSTO (the Public ESO Spectroscopic Survey for Transient Objects), ESO program ID 188.D-3003.

We thank Stefano Benetti and Massimo Della Valle for helpful comments.  We also thank Ben Shappee for kindly providing spectra of SN~2011fe.

Some of the data presented herein were obtained at the W.M. Keck Observatory, which is operated as a scientific partnership among the California Institute of Technology, the University of California and the National Aeronautics and Space Administration. The Observatory was made possible by the generous financial support of the W.M. Keck Foundation.  The authors wish to recognize and acknowledge the very significant cultural role and reverence that the summit of Mauna Kea has always had within the indigenous Hawaiian community.  We are most fortunate to have the opportunity to conduct observations from this mountain.  This research has made use of the Keck Observatory Archive (KOA), which is operated by the W. M. Keck Observatory and the NASA Exoplanet Science Institute (NExScI), under contract with the National Aeronautics and Space Administration.

This research has made use of the NASA/IPAC Extragalactic Database (NED) which is operated by the Jet Propulsion Laboratory, California Institute of Technology, under contract with the National Aeronautics and Space Administration.
This research has made use of NASA's Astrophysics Data System (ADS).

\bibliographystyle{apj}
\bibliography{nebular_snia_spectra}

\appendix

\section{Atomic Data for \CoIII}
\label{app:co_atomic_data}
{\em Rest wavelength:} Table~\ref{tab:co_atomic_data} lists the transitions in the multiplet of \CoIII\ contributing to the 5893~\AA\ feature.  Note that the second and third transitions contribute to the main 5893~\AA, while other transitions in the multiplet produce other features of interest such as the $\sim6150$~\AA\ feature of \CoIII\ (see Figure~\ref{fig:cmfgen_models}). 

\begin{table}
\renewcommand*{\arraystretch}{1.1}
\caption{Atomic data for Co\,{\sc iii}}
\centering
\begin{tabular}{ccccc}
\hline\hline
Lower level                    & Upper level                 &          A$^a$       &      $\lambda$   &        E(eV) \\
\hline
3d$^7$ a$^4$F$_{9/2}$  & 3d$^7$ a$^2$G$_{7/2}$     &    0.014    &      5627.104    &  2.203 \\
3d$^7$ a$^4$F$_{9/2}$  & 3d$^7$ a$^2$G$_{9/2}$     &    0.400    &      5888.482    &  2.105 \\
3d$^7$ a$^4$F$_{7/2}$  & 3d$^7$ a$^2$G$_{7/2}$     &    0.150    &      5906.783    &  2.203 \\
3d$^7$ a$^4$F$_{7/2}$  & 3d$^7$ a$^2$G$_{9/2}$     &    0.120    &      6195.455    &  2.105 \\
3d$^7$ a$^4$F$_{5/2}$  & 3d$^7$ a$^2$G$_{7/2}$     &    0.110    &      6127.670    &  2.203 \\
\hline
\end{tabular}
\label{tab:co_atomic_data}
\begin{flushleft}
$^a$ The $A$ values are from \citet{hansen84}
\end{flushleft}
\end{table}

\begin{figure}
\begin{center}
\includegraphics[width=0.45\textwidth]{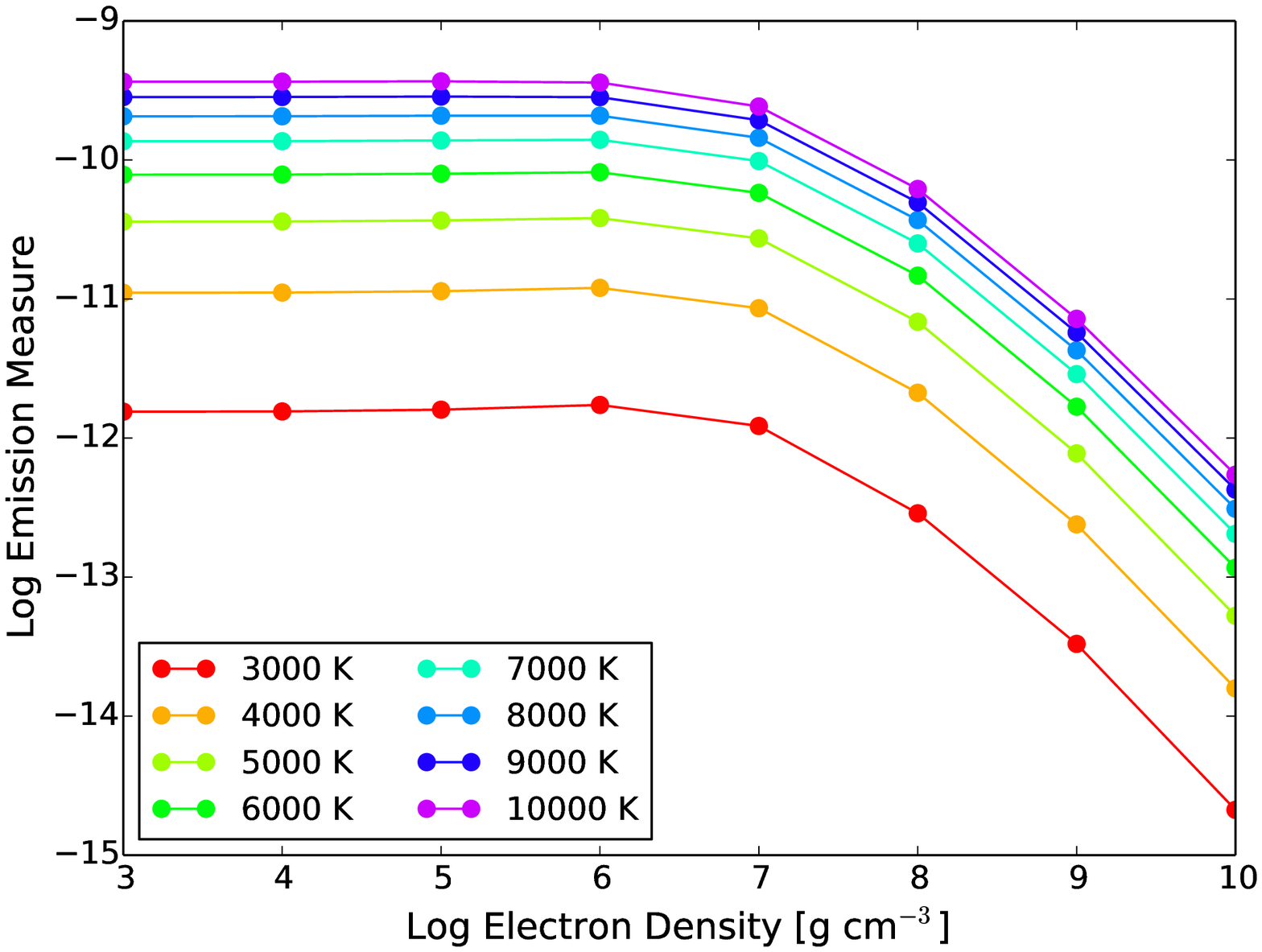}
\caption{Illustration of the variation of the flux in the \coline\ blend with electron density. The different curves are for temperatures of 3000\,K (red curve) to 10000\,K (top curve) in steps of 1000\,K. For the calculation a SN age of 100 days and a N(CoIII) to N(Ne) ratio of 0.01 was assumed. The emission measure has been divided by the product of the electron and Co densities.}
\label{fig:co_line_emissivity}
\end{center}
\end{figure}

{\em Line emissivity:}
The \coline\ blend accounts for about 70\% of the multiplet cooling. Above an electron density of $10^7$ cm$^{-3}$ and a temperature of 4000\,K, the above multiplet accounts for between 50 and 70\% of all cooling by \CoIII. As the temperature and/or electron density is lowered below these values, cooling via the 3d$^7$ a$^4$F-- 3d$^7$ a$^4$P multiplet (with an excitation energy of $\sim$1.9\,eV) becomes increasingly important.  Plots of the flux in the \coline\ blend as a function of electron density and temperature are shown in Fig.~\ref{fig:co_line_emissivity}. The critical density for the 3d$^7$ a$^2$G levels is $\sim 3 \times 10^7$ and for the 3d$^7$ a$^4$P levels it is  $\sim 1 \times 10^6$. We note here that collision rates for \CoIII\ are unavailable and so we adopted collision strengths from \cite{SNB02_NiIV_col} which were computed for Ni\,{\sc iv}, and which has a similar electronic structure to \CoIII.

\section{Ancillary data on SN sample}
\label{app:snia_data}
This Appendix presents the properties of \sneia\ and their host galaxies derived in the literature or measured from the data for our analysis.

Table~\ref{tab:host_info} presents the host galaxy distance and redshift information for the \sneia\ in our sample, along with references.  As noted in Section~\ref{sec:neb_line_fluxes}, distances for \sneia\ in our sample are preferentially derived from Cepheid distances from \citet{riess11} where available, and from host redshift with the \citet{riess11} Hubble constant value ($H_0 = 73.8$~km~s$^{-1}$~Mpc$^{-1}$).  Generally we avoid other distance indicators such as Tully-Fisher, surface brightness fluctuations, and planetary nebula luminosity functions.  The one exception to this is SN~2012cg, for which we use the \citet{cortes08} Tully-Fisher distance modulus, as was done in \citet{silverman12cg}.

For some \sneia\ (SN~2003du, SN~2011by, SN~2013dy), previous analyses in the literature have found that the redshift-based distance modulus yields anomalous values for the SN peak luminosity.  Thus for these we employ SN-based distance moduli adopted by previous authors: the \citet{stritzinger06b} value for SN~2003du, the \citet{graham15} value for SN~2011by, and the \citet{pan15} value for SN~2013dy (derived with SNooPy).

\begin{table*}
\begin{center}
\caption{Nebular SN Ia Host Information}
\label{tab:host_info}
\begin{tabular}{lllllll}
\hline
SN     & Host  & Host $z$ & Host $\mu$ & $\mu$  & $z$ Ref. & $\mu$ Ref. \\
       & Name  &          & (mag) $^a$ & Method &          &            \\
\hline
SN1990N   & NGC 4639       & $0.003395$ & $31.67 \pm 0.07$ & Cepheids     & \citet{wong06}       & \citet{riess11} \\
SN1991T   & NGC 4527       & $0.005791$ & $30.72 \pm 0.13$ & Cepheids     & \citet{zNGC4527}     & NED \\
SN1994D   & NGC 4526       & $0.002058$ & $29.73 \pm 1.06$ & Redshift     & \citet{cappellari11} & \nodata \\
SN1994ae  & NGC 3370       & $0.004266$ & $32.13 \pm 0.07$ & Cepheids     & \citet{zNGC3370}     & \citet{riess11} \\
SN1995D   & NGC 2962       & $0.006561$ & $32.25 \pm 0.33$ & Redshift     & \citet{cappellari11} & \nodata \\
SN1998aq  & NGC 3982       & $0.003699$ & $31.70 \pm 0.08$ & Cepheids     & \citet{rc3}          & \citet{riess11} \\
SN1998bu  & NGC 3368       & $0.002992$ & $30.20 \pm 0.20$ & Cepheids     & \citet{rc3}          & NED \\
SN1999aa  & NGC 2595       & $0.014443$ & $33.98 \pm 0.15$ & Redshift     & \citet{rc3}          & \nodata \\
SN2002cs  & NGC 6702       & $0.015771$ & $34.17 \pm 0.14$ & Redshift     & \citet{zNGC6702}     & \nodata \\
SN2002dj  & NGC 5018       & $0.009393$ & $33.04 \pm 0.23$ & Redshift     & \citet{zNGC5018}     & \nodata \\
SN2002er  & UGC 10743      & $0.008569$ & $32.84 \pm 0.25$ & Redshift     & \citet{rc3}          & \nodata \\
SN2002fk  & NGC 1309       & $0.007125$ & $32.59 \pm 0.09$ & Cepheids     & \citet{koribalski04} & \citet{riess11} \\
SN2003du  & UGC 9391       & $0.006384$ & $32.75 \pm 0.20$ & SN           & \citet{schneider92}  & \citet{stritzinger06b} \\
SN2003hv  & NGC 1201       & $0.005624$ & $31.92 \pm 0.39$ & Redshift     & \citet{ogando08}     & \nodata \\
SN2004bv  & NGC 6907       & $0.010614$ & $33.31 \pm 0.20$ & Redshift     & \citet{zNGC6907}     & \nodata \\
SN2004eo  & NGC 6928       & $0.015701$ & $34.16 \pm 0.14$ & Redshift     & \citet{theureau98}   & \nodata \\
SN2005am  & NGC 2811       & $0.007899$ & $32.66 \pm 0.28$ & Redshift     & \citet{theureau98}   & \nodata \\
SN2005cf  & MCG-01-39-3    & $0.006430$ & $32.21 \pm 0.34$ & Redshift     & \citet{childress13a} & \nodata \\
SN2007af  & NGC 5584       & $0.005464$ & $31.72 \pm 0.07$ & Cepheids     & \citet{koribalski04} & \citet{riess11} \\
SN2007gi  & NGC 4036       & $0.004620$ & $31.49 \pm 0.47$ & Redshift     & \citet{cappellari11} & \nodata \\
SN2007le  & NGC 7721       & $0.006721$ & $32.31 \pm 0.32$ & Redshift     & \citet{koribalski04} & \nodata \\
SN2007sr  & NGC 4038       & $0.005477$ & $31.66 \pm 0.08$ & Cepheids     & \citet{lauberts89}   & \citet{riess11} \\
SN2008Q   & NGC 524        & $0.008016$ & $32.69 \pm 0.27$ & Redshift     & \citet{cappellari11} & \nodata \\
SN2009ig  & NGC 1015       & $0.008770$ & $32.89 \pm 0.25$ & Redshift     & \citet{wong06}       & \nodata \\
SN2009le  & MCG-04-06-9    & $0.017786$ & $34.44 \pm 0.12$ & Redshift     & \citet{theureau98}   & \nodata \\
SN2011by  & NGC 3972       & $0.002843$ & $32.01 \pm 0.07$ & SN           & \citet{zNGC3972}     & \citet{graham15} \\
SN2011ek  & NGC 918        & $0.005027$ & $31.67 \pm 0.43$ & Redshift     & \citet{zNGC918}      & \nodata \\
SN2011fe  & NGC 5457       & $0.000804$ & $29.05 \pm 0.06$ & Cepheids     & \citet{rc3}          & \citet{shappee11} \\
SN2011iv  & NGC 1404       & $0.006494$ & $32.23 \pm 0.33$ & Redshift     & \citet{zNGC1404}     & \nodata \\
SN2012cg  & NGC 4424       & $0.001458$ & $30.90 \pm 0.30$ & TF           & \citet{zNGC4424}     & \citet{cortes08} \\
SN2012fr  & NGC 1365       & $0.005457$ & $31.31 \pm 0.20$ & Cepheids     & \citet{zNGC1365}     & \citet{silbermann99} \\
SN2012hr  & ESO 121-G026   & $0.007562$ & $32.56 \pm 0.29$ & Redshift     & \citet{koribalski04} & \nodata \\
SN2012ht  & NGC 3447       & $0.003559$ & $30.92 \pm 0.61$ & Redshift     & \citet{zNGC3447}     & \nodata \\
SN2013aa  & NGC 5643       & $0.003999$ & $31.18 \pm 0.54$ & Redshift     & \citet{koribalski04} & \nodata \\
SN2013cs  & ESO 576-G017   & $0.009243$ & $33.00 \pm 0.24$ & Redshift     & \citet{zSN2013cs}    & \nodata \\
SN2013dy  & NGC 7250       & $0.003889$ & $31.49 \pm 0.10$ & SN           & \citet{schneider92}  & \citet{pan15} \\
SN2013gy  & NGC 1418       & $0.014023$ & $33.92 \pm 0.15$ & Redshift     & \citet{zNGC1418}     & \nodata \\
SN2014J   & NGC 3034       & $0.000677$ & $27.60 \pm 0.10$ & Cepheids+SN  & \citet{rc3}          & \citet{foley14} \\
\hline
\end{tabular}
\end{center}
$^a$ For hosts with redshift-based $\mu$, uncertainty includes a peculiar velocity term of 300 km\,s$^{-1}$.  \\
\end{table*}

Table~\ref{tab:lc_info} presents the pertinent light curve fit parameters from the main photospheric phase light curves.  Light curves were fit with SiFTO \citep{sifto}.  For some \sneia\ (SN~2011iv, SN~2011by, SN~2012ht), photometry was not explicitly published, so we utilize the published light curve widths and colours for these SNe.  Where appropriate, we convert \dmft\ or SALT2 $x1$ to SiFTO stretch $s$ using the relations published in \citet{guy07, guy10}.

Light curves for SN~2013cs and SN~2013gy were obtained with the Las Cumbres Observatories Global Telescope \citep[LCOGT][]{lcogt} and have been reduced using a custom pipeline developed by S. Valenti. The pipeline employs standard procedures (PyRAF, DAOPHOT) in a Python framework. Host galaxy flux was removed using a low order polynomial background. Point spread function magnitudes were transformed to the standard Sloan Digital Sky Survey \citep{smith02} filter system (for gri) or \citet{landolt92} system (for BV) via standard star observations taken during clear nights.  These light curves will be released in a future LCOGT supernova program publication.

\begin{table*}
\begin{center}
\caption{Light Curve Fit Results and Sources}
\label{tab:lc_info}
\begin{tabular}{llllll}
\hline
SN     & MJD         & Rest Frame       & Stretch   & Color $c$ & LC \\
       & of $B_{max}$ & $B_{max}$ (mag) &           &           & Ref.  \\
\hline
SN1990N   & $48081.69$ & $12.706 \pm 0.015$ & $1.081 \pm 0.012$ & $ 0.049 \pm 0.013$ & \citet{lira98} \\
SN1991T   & $48373.95$ & $11.468 \pm 0.028$ & $1.047 \pm 0.019$ & $ 0.101 \pm 0.026$ & \citet{lira98} \\
SN1994D   & $49431.52$ & $11.775 \pm 0.009$ & $0.823 \pm 0.004$ & $-0.099 \pm 0.005$ & \citet{patat96} \\
SN1994ae  & $49684.56$ & $12.968 \pm 0.019$ & $1.057 \pm 0.014$ & $-0.060 \pm 0.016$ & CfA \\
SN1995D   & $49767.50$ & $13.273 \pm 0.025$ & $1.097 \pm 0.014$ & $-0.006 \pm 0.015$ & CfA \\
SN1998aq  & $50930.24$ & $12.316 \pm 0.009$ & $0.965 \pm 0.010$ & $-0.146 \pm 0.009$ & \citet{riess05} \\
SN1998bu  & $50952.06$ & $12.118 \pm 0.013$ & $0.962 \pm 0.018$ & $ 0.270 \pm 0.010$ & \citet{jha99, suntzeff99} \\
SN1999aa  & $51231.84$ & $14.755 \pm 0.016$ & $1.134 \pm 0.009$ & $-0.047 \pm 0.009$ & CfA \\
SN2002cs  & $52409.24$ & $15.138 \pm 0.039$ & $1.007 \pm 0.013$ & $ 0.017 \pm 0.016$ & LOSS \\
SN2002dj  & $52450.32$ & $13.974 \pm 0.035$ & $0.962 \pm 0.013$ & $ 0.098 \pm 0.016$ & LOSS \\
SN2002er  & $52524.03$ & $14.267 \pm 0.057$ & $0.901 \pm 0.009$ & $ 0.123 \pm 0.018$ & LOSS \\
SN2002fk  & $52547.28$ & $13.152 \pm 0.017$ & $0.995 \pm 0.012$ & $-0.142 \pm 0.012$ & LOSS \\
SN2003du  & $52765.48$ & $13.476 \pm 0.007$ & $1.016 \pm 0.008$ & $-0.110 \pm 0.008$ & LOSS \\
SN2003hv  & $52891.12$ & $12.444 \pm 0.020$ & $0.741 \pm 0.021$ & $-0.115 \pm 0.014$ & \citet{leloudas09} \\
SN2004bv  & $53159.83$ & $13.938 \pm 0.024$ & $1.146 \pm 0.014$ & $ 0.122 \pm 0.013$ & LOSS \\
SN2004eo  & $53277.66$ & $15.099 \pm 0.038$ & $0.863 \pm 0.009$ & $ 0.002 \pm 0.014$ & CSP \\
SN2005am  & $53436.97$ & $13.698 \pm 0.023$ & $0.710 \pm 0.019$ & $ 0.057 \pm 0.012$ & CSP \\
SN2005cf  & $53533.28$ & $13.625 \pm 0.035$ & $0.968 \pm 0.009$ & $ 0.021 \pm 0.013$ & LOSS \\
SN2007af  & $54173.88$ & $13.180 \pm 0.015$ & $0.955 \pm 0.011$ & $ 0.058 \pm 0.010$ & CSP \\
SN2007gi  & $54327.66$ & $13.158 \pm 0.016$ & $0.871 \pm 0.013$ & $ 0.097 \pm 0.017$ & \citet{zhang10} \\
SN2007le  & $54398.62$ & $13.876 \pm 0.016$ & $1.028 \pm 0.015$ & $ 0.342 \pm 0.014$ & CSP \\
SN2007sr  & $54448.70$ & $12.809 \pm 0.042$ & $0.970 \pm 0.031$ & $ 0.141 \pm 0.016$ & LOSS \\
SN2008Q   & $54505.47$ & $13.510 \pm 0.033$ & $0.803 \pm 0.025$ & $ 0.001 \pm 0.018$ & LOSS \\
SN2009ig  & $55079.18$ & $13.437 \pm 0.019$ & $1.136 \pm 0.028$ & $ 0.055 \pm 0.019$ & CfA \\
SN2009le  & $55165.16$ & $15.351 \pm 0.030$ & $1.022 \pm 0.022$ & $ 0.079 \pm 0.033$ & CfA \\
SN2011by  & $55690.60$ & $12.890 \pm 0.030$ & $0.959 \pm 0.019$ & $ 0.000 \pm 0.050$ & \citet{graham15} \\
SN2011ek  & $55788.90$ & $13.840 \pm 0.150$ & $0.900 \pm 0.020$ & $ 0.180 \pm 0.040$ & \citet{maguire13} \\
SN2011fe  & $55814.51$ & $ 9.940 \pm 0.010$ & $0.969 \pm 0.010$ & $-0.066 \pm 0.021$ & \citet{pereira13} \\
SN2011iv  & $55906.00$ & $12.530 \pm 0.040$ & $0.684 \pm 0.020$ & $ 0.000 \pm 0.050$ & \citet{foley12b} \\
SN2012cg  & $56082.03$ & $12.128 \pm 0.011$ & $1.063 \pm 0.011$ & $ 0.184 \pm 0.010$ & \citet{munari13} \\
SN2012fr  & $56243.68$ & $12.017 \pm 0.013$ & $1.108 \pm 0.008$ & $ 0.059 \pm 0.011$ & \citet{zhang14} \\
SN2012hr  & $56289.20$ & $13.780 \pm 0.020$ & $1.018 \pm 0.021$ & $ 0.030 \pm 0.010$ & \citet{maguire13} \\
SN2012ht  & $56295.60$ & $13.220 \pm 0.040$ & $0.818 \pm 0.025$ & $ 0.000 \pm 0.050$ & \citet{yamanaka14} \\
SN2013aa  & $56344.00$ & $11.330 \pm 0.050$ & $1.146 \pm 0.019$ & $-0.050 \pm 0.010$ & \citet{maguire13} \\
SN2013cs  & $56436.95$ & $13.659 \pm 0.032$ & $1.006 \pm 0.010$ & $ 0.054 \pm 0.015$ & LCOGT \\
SN2013dy  & $56500.65$ & $12.824 \pm 0.048$ & $1.101 \pm 0.007$ & $ 0.195 \pm 0.015$ & \citet{pan15b} \\
SN2013gy  & $56648.61$ & $14.742 \pm 0.020$ & $0.948 \pm 0.008$ & $ 0.070 \pm 0.011$ & LCOGT \\
SN2014J   & $56689.74$ & $11.850 \pm 0.012$ & $1.086 \pm 0.010$ & $ 1.251 \pm 0.012$ & \citet{foley14} \\
\hline
\end{tabular}
\end{center}
CfA: \citet{riess99, jha06, hicken09a, hicken12}; LOSS: \citet{ganesh10}; CSP: \citet{contreras10, stritzinger11}
\end{table*}

Table~\ref{tab:late_phot} presents the late-time photometry for our \sneia.  The $B$-band magnitudes for each phase of spectroscopic observation was determined from a linear fit to the very late ($t \geq 60$~days past peak) $B$-band light curve.  Spectra from T15b were already flux calibrated to contemporaneous photometry, so here we report the $B$-band flux synthesized from the spectrum.  Similarly, M15 spectra were flux calibrated with spectrophotometric standard stars, and thus are flux calibrated modulo any changes in grey extinction or slit loss between the standard star and SN observation.  

\begin{figure*}
\begin{center}
\includegraphics[width=0.90\textwidth]{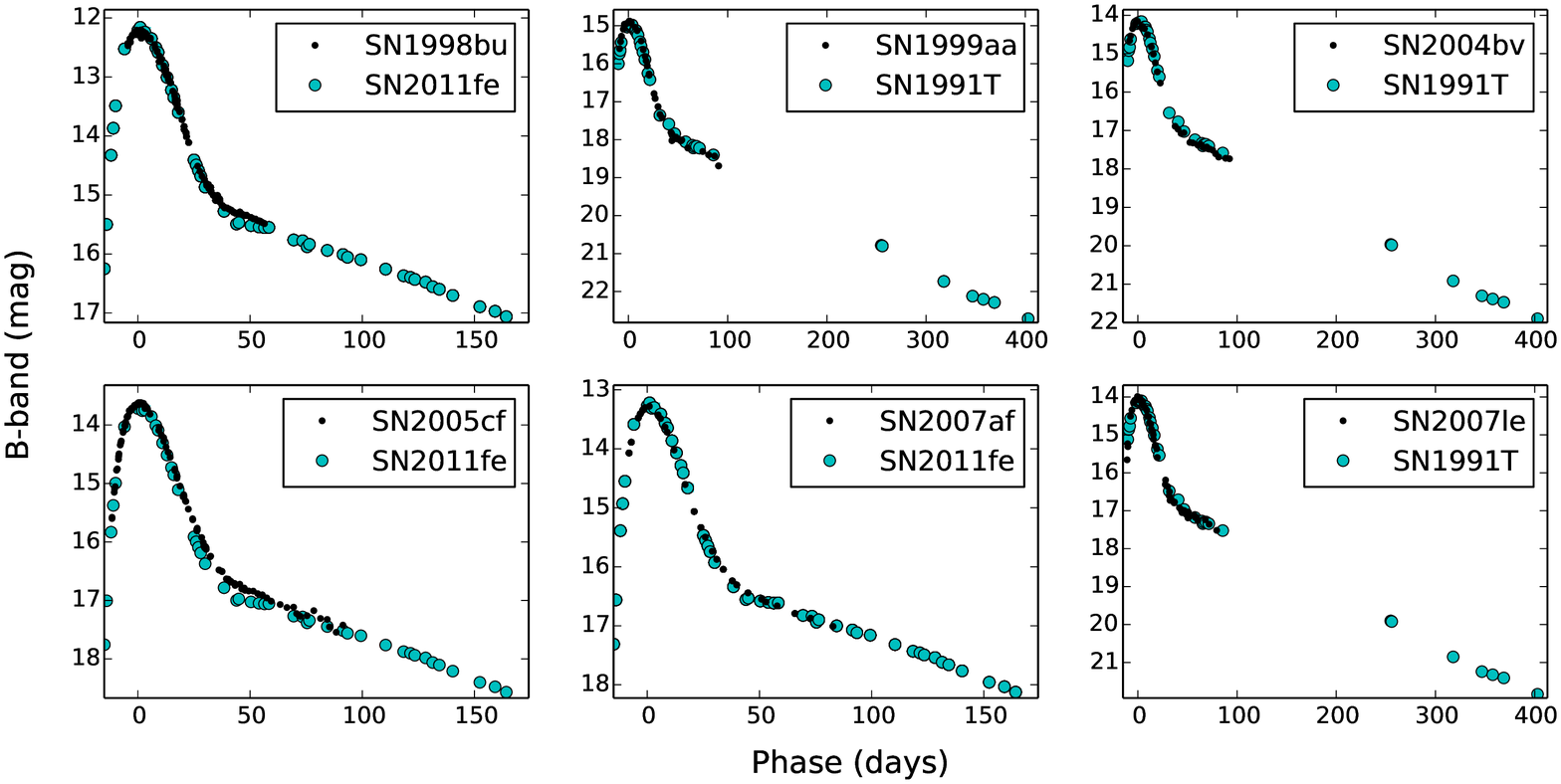}
\caption{Surrogate light curves for \sneia\ with poor late time light curve coverage.  Original SN data are shown as small black dots while surrogate SN data are shown as larger blue circles.}
\label{fig:surrogate_late_lcs}
\end{center}
\end{figure*}

For some \sneia, sufficient late time photometry was not available to reliably extrapolate the $B$-band luminosity at the time of spectroscopic observations -- generally this occurred if the SN did not have photometry beyond 100 days past peak or if a public light curve was not available.  For these SNe, late photometry from an alternate SN was employed, with the requirement that the light curve stretch be exceptionally close.  These SNe with their surrogate light curves are shown in Figure~\ref{fig:surrogate_late_lcs}, and the surrogate SN is denoted in the reference column of Table~\ref{tab:late_phot}.

Table~\ref{tab:line_fit_results} presents the measured \coline\ line fluxes, synthesized $B$-band fluxes, and wavelength integration bounds for the \coline\ flux measurements.  Note that the fluxes in this table are in the natural units of the spectroscopic data, as measured from its publicly available format.

\clearpage
\begin{deluxetable}{lllll}
\tablecaption{Late Phase Photometry
\label{tab:late_phot}}
\tablewidth{0pt}
\tabletypesize{\small}
\tablehead{
\colhead{SN} &
\colhead{Phase $t$} &
\colhead{Obs. Date} &
\colhead{$m_B(t)$} &
\colhead{Phot. Ref.} \\
\colhead{} &
\colhead{(days)} &
\colhead{} &
\colhead{ (mag)} &
\colhead{}
}
\startdata
SN1990N   & 160 & 19901217 & $17.394 \pm 0.014$ & \citet{lira98} \\
SN1990N   & 186 & 19910112 & $17.760 \pm 0.011$ &  \\
SN1990N   & 227 & 19910222 & $18.337 \pm 0.007$ &  \\
SN1990N   & 255 & 19910322 & $18.731 \pm 0.008$ &  \\
SN1990N   & 280 & 19910416 & $19.083 \pm 0.011$ &  \\
SN1990N   & 333 & 19910608 & $19.829 \pm 0.018$ &  \\
SN1991T   & 113 & 19910819 & $17.026 \pm 0.032$ & \citet{schmidt94} \\
SN1991T   & 186 & 19911031 & $17.614 \pm 0.024$ &  \\
SN1991T   & 258 & 19920111 & $18.194 \pm 0.018$ &  \\
SN1991T   & 320 & 19920313 & $18.694 \pm 0.018$ &  \\
SN1991T   & 349 & 19920411 & $18.928 \pm 0.019$ &  \\
SN1993Z   & 173 & 19940317 & $19.064 \pm 0.063$ & \citet{ho01} \\
SN1993Z   & 205 & 19940418 & $19.588 \pm 0.094$ &  \\
SN1994D   & 613 & 19951124 & $23.227 \pm 0.040$ & \citet{altavilla04} \\
SN1994ae  & 144 & 19950422 & $17.672 \pm 0.025$ & \citet{altavilla04} \\
SN1994ae  & 153 & 19950501 & $17.812 \pm 0.027$ &  \\
SN1995D   & 277 & 19951124 & $19.887 \pm 0.041$ & \citet{altavilla04} \\
SN1995D   & 285 & 19951202 & $20.001 \pm 0.043$ &  \\
SN1996X   & 298 & 19970210 & $20.108 \pm 0.022$ & \citet{salvo01} \\
SN1998aq  & 211 & 19981124 & $18.140 \pm 0.011$ & \citet{riess05} \\
SN1998aq  & 231 & 19981214 & $18.442 \pm 0.013$ &  \\
SN1998aq  & 241 & 19981224 & $18.593 \pm 0.013$ &  \\
SN1998bu  & 179 & 19981114 & $17.202 \pm 0.017$ & {\bf SN2011fe} \\
SN1998bu  & 190 & 19981125 & $17.347 \pm 0.019$ &  \\
SN1998bu  & 208 & 19981213 & $17.586 \pm 0.022$ &  \\
SN1998bu  & 217 & 19981222 & $17.705 \pm 0.023$ &  \\
SN1998bu  & 236 & 19990110 & $17.956 \pm 0.026$ &  \\
SN1998bu  & 243 & 19990117 & $18.049 \pm 0.028$ &  \\
SN1998bu  & 280 & 19990223 & $18.538 \pm 0.034$ &  \\
SN1998bu  & 329 & 19990413 & $19.187 \pm 0.042$ &  \\
SN1998bu  & 340 & 19990424 & $19.332 \pm 0.043$ &  \\
SN1999aa  & 256 & 19991109 & $20.872 \pm 0.000$ & {\bf SN1991T} \\
SN1999aa  & 282 & 19991205 & $21.236 \pm 0.000$ &  \\
SN2002cs  & 174 & 20021106 & $20.264 \pm 0.080$ & LOSS \\
SN2002dj  & 222 & 20030201 & $20.143 \pm 0.058$ & \citet{pignata04} \\
SN2002dj  & 275 & 20030326 & $21.003 \pm 0.040$ &  \\
SN2002er  & 216 & 20030410 & $20.429 \pm 0.022$ & \citet{pignata08} \\
SN2002fk  & 150 & 20030227 & $18.099 \pm 0.024$ & CfA, LOSS \\
SN2003du  & 109 & 20030823 & $17.612 \pm 0.004$ & \citet{stanishev07} \\
SN2003du  & 138 & 20030921 & $18.043 \pm 0.003$ &  \\
SN2003du  & 139 & 20030922 & $18.058 \pm 0.003$ &  \\
SN2003du  & 142 & 20030925 & $18.102 \pm 0.003$ &  \\
SN2003du  & 209 & 20031201 & $19.096 \pm 0.004$ &  \\
SN2003du  & 221 & 20031213 & $19.274 \pm 0.004$ &  \\
SN2003du  & 272 & 20040202 & $20.031 \pm 0.005$ &  \\
SN2003du  & 377 & 20040517 & $21.589 \pm 0.008$ &  \\
SN2003gs  & 170 & 20040214 & $19.877 \pm 0.051$ & LOSS \\
SN2003hv  & 113 & 20031228 & $16.927 \pm 0.003$ & \citet{leloudas09} \\
SN2003hv  & 145 & 20040129 & $17.436 \pm 0.004$ &  \\
SN2003hv  & 323 & 20040725 & $20.870 \pm 0.012$ &  \\
SN2004bv  & 171 & 20041114 & $18.713 \pm 0.000$ & {\bf SN1991T} \\
SN2004eo  & 228 & 20050516 & $21.454 \pm 0.008$ & \citet{pastorello07} \\
SN2005cf  & 319 & 20060427 & $20.558 \pm 0.040$ & {\bf SN2011fe} \\
SN2006X   & 127 & 20060626 & $19.001 \pm 0.027$ & \citet{wang08} \\
SN2006X   & 152 & 20060721 & $19.196 \pm 0.040$ &  \\
SN2006X   & 277 & 20061123 & $20.175 \pm 0.107$ &  \\
SN2006X   & 307 & 20061223 & $20.410 \pm 0.123$ &  \\
SN2006X   & 360 & 20070214 & $20.824 \pm 0.151$ &  \\
SN2007af  & 103 & 20070620 & $17.181 \pm 0.005$ & {\bf SN2011fe} \\
SN2007af  & 108 & 20070625 & $17.247 \pm 0.006$ &  \\
SN2007af  & 120 & 20070707 & $17.406 \pm 0.007$ &  \\
SN2007af  & 123 & 20070710 & $17.446 \pm 0.008$ &  \\
SN2007af  & 128 & 20070715 & $17.512 \pm 0.008$ &  \\
SN2007af  & 131 & 20070718 & $17.552 \pm 0.009$ &  \\
SN2007af  & 151 & 20070807 & $17.816 \pm 0.012$ &  \\
SN2007af  & 165 & 20070821 & $18.002 \pm 0.014$ &  \\
SN2007af  & 308 & 20080111 & $19.894 \pm 0.037$ &  \\
SN2007gi  & 161 & 20080115 & $17.920 \pm 0.014$ & \citet{zhang10} \\
SN2007le  & 317 & 20080827 & $20.653 \pm 0.000$ & {\bf SN1991T} \\
SN2007sr  & 177 & 20080623 & $17.994 \pm 0.021$ & CfA, LOSS \\
SN2009ig  & 405 & 20101017 & $21.729 \pm 0.050$ & CfA \\
SN2009le  & 324 & 20101016 & $22.651 \pm 0.050$ & CfA \\
SN2010ev  & 178 & 20101231 & $20.199 \pm 0.050$ & T15b \\
SN2010ev  & 272 & 20110404 & $21.183 \pm 0.050$ &  \\
SN2010gp  & 279 & 20110501 & $23.150 \pm 0.050$ & T15b \\
SN2010hg  & 199 & 20110403 & $19.765 \pm 0.050$ & T15b \\
SN2010kg  & 289 & 20110927 & $22.592 \pm 0.050$ & T15b \\
SN2011K   & 341 & 20111228 & $23.096 \pm 0.050$ & T15b \\
SN2011ae  & 310 & 20120101 & $20.671 \pm 0.050$ & T15b \\
SN2011at  & 349 & 20120227 & $21.656 \pm 0.050$ & T15b \\
SN2011by  & 206 & 20111202 & $18.337 \pm 0.022$ & {\bf SN2011fe} \\
SN2011by  & 310 & 20120315 & $19.713 \pm 0.039$ &  \\
SN2011ek  & 423 & 20121011 & $24.328 \pm 0.050$ & T15b \\
SN2011fe  & 074 & 20111123 & $13.630 \pm 0.004$ & \citet{richmond12} \\
SN2011fe  & 114 & 20120102 & $14.157 \pm 0.007$ &  \\
SN2011fe  & 196 & 20120324 & $15.238 \pm 0.019$ &  \\
SN2011fe  & 230 & 20120427 & $15.686 \pm 0.024$ &  \\
SN2011fe  & 276 & 20120612 & $16.292 \pm 0.030$ &  \\
SN2011fe  & 314 & 20120720 & $16.792 \pm 0.036$ &  \\
SN2011im  & 314 & 20121016 & $23.522 \pm 0.050$ & T15b \\
SN2011iv  & 318 & 20121024 & $20.717 \pm 0.050$ & T15b \\
SN2011jh  & 414 & 20130221 & $24.131 \pm 0.050$ & T15b \\
SN2012cg  & 330 & 20130507 & $19.824 \pm 0.085$ & {\bf SN1994ae} \\
SN2012cg  & 342 & 20130513 & $19.917 \pm 0.087$ &  \\
SN2012cu  & 340 & 20130603 & $23.060 \pm 0.050$ & T15b \\
SN2012fr  & 101 & 20130221 & $15.696 \pm 0.015$ & \citet{zhang14} \\
SN2012fr  & 116 & 20130308 & $15.927 \pm 0.020$ &  \\
SN2012fr  & 125 & 20130317 & $16.065 \pm 0.024$ &  \\
SN2012fr  & 151 & 20130412 & $16.466 \pm 0.034$ &  \\
SN2012fr  & 222 & 20130622 & $17.560 \pm 0.065$ &  \\
SN2012fr  & 261 & 20130731 & $18.161 \pm 0.082$ &  \\
SN2012fr  & 340 & 20131018 & $19.379 \pm 0.117$ &  \\
SN2012fr  & 357 & 20131103 & $19.409 \pm 0.100$ &  \\
SN2012fr  & 357 & 20131103 & $19.625 \pm 0.124$ &  \\
SN2012fr  & 367 & 20131114 & $19.795 \pm 0.129$ &  \\
SN2012hr  & 283 & 20131006 & $20.487 \pm 0.005$ & {\bf SN2003du} \\
SN2012hr  & 368 & 20131230 & $21.749 \pm 0.007$ &  \\
SN2012ht  & 437 & 20140313 & $22.714 \pm 0.100$ & T15b \\
SN2013aa  & 137 & 20130710 & $15.768 \pm 0.000$ & {\bf SN1991T} \\
SN2013aa  & 185 & 20130827 & $16.438 \pm 0.000$ &  \\
SN2013aa  & 202 & 20130913 & $16.676 \pm 0.000$ &  \\
SN2013aa  & 342 & 20140131 & $18.633 \pm 0.001$ &  \\
SN2013aa  & 358 & 20140216 & $18.857 \pm 0.001$ &  \\
SN2013aa  & 430 & 20140422 & $19.765 \pm 0.001$ &  \\
SN2013cs  & 320 & 20140322 & $20.909 \pm 0.068$ & {\bf SN2002fk} \\
SN2013cs  & 322 & 20140324 & $20.939 \pm 0.069$ &  \\
SN2013cs  & 322 & 20140324 & $21.475 \pm 0.100$ &  \\
SN2013ct  & 223 & 20131218 & $18.144 \pm 0.100$ & T15b \\
SN2013dl  & 160 & 20141130 & $20.034 \pm 0.100$ & T15b \\
SN2013dl  & 184 & 20141224 & $20.331 \pm 0.100$ &  \\
SN2013dy  & 333 & 20140626 & $20.716 \pm 0.033$ & \citet{pan15} \\
SN2013dy  & 419 & 20140920 & $22.064 \pm 0.045$ &  \\
SN2013ef  & 174 & 20141224 & $21.511 \pm 0.100$ & T15b \\
SN2013gy  & 276 & 20140920 & $21.242 \pm 0.062$ & {\bf SN2007af} \\
SN2014J   & 231 & 20140920 & $16.956 \pm 0.119$ & \citet{foley14} \\
\enddata
\tablecomments{
For SNe with poor late-time photometric coverage, the late light curve of a surrogate SN (denoted in {\bf boldface} in the Photometry Reference column) is employed.\\
CfA: \citet{riess99, jha06, hicken09a, hicken12}; LOSS: \citet{ganesh10}; CSP: \citet{contreras10, stritzinger11}
}
\end{deluxetable}

\clearpage
\begin{deluxetable}{llllll}
\tablecaption{Co line fluxes from spectra
\label{tab:line_fit_results}}
\tablewidth{0pt}
\tabletypesize{\small}
\tablehead{
\colhead{SN} &
\colhead{Date} &
\colhead{Co Line Flux} &
\colhead{B-Band Flux} &
\colhead{$w_{min}$} &
\colhead{$w_{max}$} \\
\colhead{} &
\colhead{} &
\colhead{(erg\,cm$^2$\,s$^{-1}$)} &
\colhead{(erg\,cm$^2$\,s$^{-1}$)} &
\colhead{(\AA)} &
\colhead{(\AA)}
}
\startdata
SN1990N   & 19901217 & $6.06\times 10^{-16} \pm 3.39\times 10^{-18}$ & $3.52\times 10^{-15} \pm 1.95\times 10^{-17}$ & 5710.5 & 6038.4 \\
SN1990N   & 19910112 & $5.99\times 10^{-14} \pm 1.76\times 10^{-17}$ & $3.98\times 10^{-13} \pm 7.37\times 10^{-17}$ &  &  \\
SN1990N   & 19910222 & $3.69\times 10^{-14} \pm 8.73\times 10^{-17}$ & $3.30\times 10^{-13} \pm 5.29\times 10^{-16}$ &  &  \\
SN1990N   & 19910322 & $9.78\times 10^{-15} \pm 4.25\times 10^{-17}$ & $1.28\times 10^{-13} \pm 2.03\times 10^{-16}$ &  &  \\
SN1990N   & 19910416 & $7.15\times 10^{-15} \pm 5.64\times 10^{-17}$ & $1.03\times 10^{-13} \pm 2.24\times 10^{-16}$ &  &  \\
SN1990N   & 19910608 & $1.89\times 10^{-15} \pm 4.67\times 10^{-17}$ & $5.34\times 10^{-14} \pm 1.46\times 10^{-16}$ &  &  \\
SN1991M   & 19910820 & $1.67\times 10^{-16} \pm 5.39\times 10^{-18}$ & $8.47\times 10^{-16} \pm 1.29\times 10^{-17}$ & 5770.8 & 6031.1 \\
SN1991T   & 19910819 & $4.99\times 10^{-14} \pm 1.98\times 10^{-16}$ & $1.28\times 10^{-13} \pm 5.48\times 10^{-16}$ & 5664.2 & 6121.0 \\
SN1991T   & 19911031 & $1.34\times 10^{-15} \pm 2.23\times 10^{-17}$ & $8.20\times 10^{-15} \pm 4.80\times 10^{-17}$ &  &  \\
SN1991T   & 19920111 & $5.49\times 10^{-14} \pm 8.12\times 10^{-17}$ & $4.57\times 10^{-13} \pm 2.40\times 10^{-16}$ &  &  \\
SN1991T   & 19920313 & $1.18\times 10^{-16} \pm 4.46\times 10^{-18}$ & $1.09\times 10^{-15} \pm 6.92\times 10^{-18}$ &  &  \\
SN1991T   & 19920411 & $5.67\times 10^{-17} \pm 2.51\times 10^{-18}$ & $8.32\times 10^{-16} \pm 3.12\times 10^{-17}$ &  &  \\
SN1993Z   & 19940317 & $2.17\times 10^{-16} \pm 3.14\times 10^{-18}$ & $1.14\times 10^{-15} \pm 1.00\times 10^{-17}$ & 5757.8 & 6081.7 \\
SN1993Z   & 19940418 & $9.90\times 10^{-17} \pm 2.47\times 10^{-18}$ & $9.73\times 10^{-16} \pm 5.87\times 10^{-18}$ &  &  \\
SN1994D   & 19951124 & $4.69\times 10^{-15} \pm 3.19\times 10^{-16}$ & $4.88\times 10^{-14} \pm 4.08\times 10^{-16}$ & 5732.7 & 6070.7 \\
SN1994ae  & 19950422 & $6.74\times 10^{-16} \pm 7.37\times 10^{-18}$ & $3.77\times 10^{-15} \pm 9.61\times 10^{-18}$ & 5747.5 & 6027.4 \\
SN1994ae  & 19950501 & $5.38\times 10^{-14} \pm 1.84\times 10^{-16}$ & $4.04\times 10^{-13} \pm 6.40\times 10^{-16}$ &  &  \\
SN1995D   & 19951124 & $4.96\times 10^{-15} \pm 3.49\times 10^{-16}$ & $4.88\times 10^{-14} \pm 4.08\times 10^{-16}$ & 5677.7 & 6080.5 \\
SN1995D   & 19951202 & $1.96\times 10^{-15} \pm 1.70\times 10^{-16}$ & $2.67\times 10^{-14} \pm 2.00\times 10^{-16}$ &  &  \\
SN1996X   & 19970210 & $2.45\times 10^{-15} \pm 3.74\times 10^{-17}$ & $2.97\times 10^{-14} \pm 1.09\times 10^{-16}$ & 5681.8 & 6075.7 \\
SN1998aq  & 19981124 & $3.80\times 10^{-14} \pm 1.20\times 10^{-15}$ & $2.68\times 10^{-13} \pm 2.37\times 10^{-15}$ & 5668.6 & 6077.6 \\
SN1998aq  & 19981214 & $5.90\times 10^{-14} \pm 4.93\times 10^{-16}$ & $3.66\times 10^{-13} \pm 1.01\times 10^{-15}$ &  &  \\
SN1998aq  & 19981224 & $4.86\times 10^{-14} \pm 7.15\times 10^{-16}$ & $3.86\times 10^{-13} \pm 1.79\times 10^{-15}$ &  &  \\
SN1998bu  & 19981114 & $1.27\times 10^{-13} \pm 7.07\times 10^{-16}$ & $5.36\times 10^{-13} \pm 1.95\times 10^{-15}$ & 5706.7 & 6055.0 \\
SN1998bu  & 19981125 & $1.39\times 10^{-13} \pm 5.29\times 10^{-16}$ & $6.45\times 10^{-13} \pm 1.20\times 10^{-15}$ &  &  \\
SN1998bu  & 19981213 & $8.76\times 10^{-14} \pm 3.94\times 10^{-16}$ & $5.13\times 10^{-13} \pm 9.64\times 10^{-16}$ &  &  \\
SN1998bu  & 19981222 & $8.04\times 10^{-14} \pm 6.98\times 10^{-16}$ & $4.24\times 10^{-13} \pm 1.59\times 10^{-15}$ &  &  \\
SN1998bu  & 19990110 & $4.58\times 10^{-16} \pm 1.19\times 10^{-18}$ & $3.71\times 10^{-15} \pm 5.44\times 10^{-18}$ &  &  \\
SN1998bu  & 19990117 & $5.47\times 10^{-14} \pm 3.47\times 10^{-16}$ & $2.95\times 10^{-13} \pm 8.72\times 10^{-16}$ &  &  \\
SN1998bu  & 19990223 & $2.08\times 10^{-16} \pm 1.23\times 10^{-18}$ & $1.72\times 10^{-15} \pm 3.75\times 10^{-18}$ &  &  \\
SN1998bu  & 19990413 & $6.52\times 10^{-15} \pm 7.18\times 10^{-17}$ & $6.01\times 10^{-14} \pm 1.06\times 10^{-16}$ &  &  \\
SN1998bu  & 19990424 & $1.42\times 10^{-16} \pm 2.54\times 10^{-18}$ & $1.04\times 10^{-15} \pm 7.50\times 10^{-18}$ &  &  \\
SN1999aa  & 19991109 & $1.80\times 10^{-16} \pm 1.12\times 10^{-18}$ & $8.69\times 10^{-16} \pm 2.46\times 10^{-18}$ & 5616.9 & 6197.1 \\
SN1999aa  & 19991205 & $7.22\times 10^{-17} \pm 8.10\times 10^{-19}$ & $3.46\times 10^{-16} \pm 2.46\times 10^{-18}$ &  &  \\
SN2002cs  & 20021106 & $1.34\times 10^{-16} \pm 6.21\times 10^{-19}$ & $6.04\times 10^{-16} \pm 2.59\times 10^{-17}$ & 5655.5 & 6097.2 \\
SN2002dj  & 20030326 & $3.41\times 10^{-15} \pm 2.40\times 10^{-17}$ & $2.31\times 10^{-14} \pm 3.97\times 10^{-17}$ & 5678.4 & 6148.5 \\
SN2002er  & 20030410 & $5.02\times 10^{-15} \pm 8.81\times 10^{-17}$ & $4.18\times 10^{-14} \pm 2.10\times 10^{-16}$ & 5709.1 & 6086.8 \\
SN2002fk  & 20030227 & $4.95\times 10^{-16} \pm 6.66\times 10^{-19}$ & $3.08\times 10^{-15} \pm 2.47\times 10^{-18}$ & 5739.0 & 6008.7 \\
SN2003du  & 20030823 & $1.36\times 10^{-13} \pm 3.14\times 10^{-16}$ & $5.08\times 10^{-13} \pm 1.06\times 10^{-15}$ & 5736.0 & 6027.0 \\
SN2003du  & 20030921 & $4.90\times 10^{-14} \pm 9.36\times 10^{-17}$ & $2.47\times 10^{-13} \pm 4.95\times 10^{-16}$ &  &  \\
SN2003du  & 20030922 & $4.24\times 10^{-14} \pm 1.16\times 10^{-16}$ & $2.12\times 10^{-13} \pm 6.54\times 10^{-16}$ &  &  \\
SN2003du  & 20030925 & $5.96\times 10^{-14} \pm 1.36\times 10^{-16}$ & $3.00\times 10^{-13} \pm 7.96\times 10^{-16}$ &  &  \\
SN2003du  & 20031201 & $1.39\times 10^{-14} \pm 1.19\times 10^{-16}$ & $1.05\times 10^{-13} \pm 3.34\times 10^{-16}$ &  &  \\
SN2003du  & 20031213 & $8.91\times 10^{-15} \pm 1.06\times 10^{-16}$ & $9.51\times 10^{-14} \pm 2.63\times 10^{-16}$ &  &  \\
SN2003du  & 20040202 & $3.34\times 10^{-15} \pm 6.14\times 10^{-17}$ & $6.68\times 10^{-14} \pm 2.01\times 10^{-16}$ &  &  \\
SN2003du  & 20040517 & $5.05\times 10^{-16} \pm 7.93\times 10^{-18}$ & $1.24\times 10^{-14} \pm 1.96\times 10^{-17}$ &  &  \\
SN2003gs  & 20040214 & $4.13\times 10^{-17} \pm 2.53\times 10^{-19}$ & $4.83\times 10^{-16} \pm 1.08\times 10^{-17}$ & 5762.4 & 6006.7 \\
SN2003hv  & 20031228 & $4.12\times 10^{-13} \pm 1.28\times 10^{-15}$ & $1.64\times 10^{-12} \pm 2.68\times 10^{-15}$ & 5687.6 & 6061.5 \\
SN2003hv  & 20040129 & $3.33\times 10^{-13} \pm 1.31\times 10^{-15}$ & $1.96\times 10^{-12} \pm 3.52\times 10^{-15}$ &  &  \\
SN2003hv  & 20040725 & $2.17\times 10^{-15} \pm 1.56\times 10^{-17}$ & $3.00\times 10^{-14} \pm 3.21\times 10^{-17}$ &  &  \\
SN2004bv  & 20041114 & $3.89\times 10^{-16} \pm 1.05\times 10^{-18}$ & $2.00\times 10^{-15} \pm 1.87\times 10^{-18}$ & 5744.3 & 6057.6 \\
SN2004eo  & 20050516 & $1.46\times 10^{-15} \pm 4.14\times 10^{-18}$ & $1.22\times 10^{-14} \pm 1.10\times 10^{-17}$ & 5737.5 & 6034.0 \\
SN2005cf  & 20060427 & $4.44\times 10^{-18} \pm 2.06\times 10^{-19}$ & $5.34\times 10^{-17} \pm 1.99\times 10^{-19}$ & 5764.7 & 5992.4 \\
SN2006X   & 20060626 & $7.25\times 10^{-16} \pm 1.12\times 10^{-17}$ & $8.76\times 10^{-16} \pm 1.13\times 10^{-17}$ & 5670.0 & 6118.6 \\
SN2006X   & 20060721 & $3.12\times 10^{-14} \pm 4.78\times 10^{-16}$ & $5.35\times 10^{-14} \pm 1.20\times 10^{-15}$ &  &  \\
SN2006X   & 20061123 & $7.25\times 10^{-17} \pm 5.51\times 10^{-19}$ & $2.01\times 10^{-16} \pm 7.62\times 10^{-19}$ &  &  \\
SN2006X   & 20061223 & $3.70\times 10^{-16} \pm 2.24\times 10^{-18}$ & $5.19\times 10^{-16} \pm 5.02\times 10^{-17}$ &  &  \\
SN2006X   & 20070214 & $5.51\times 10^{-17} \pm 1.42\times 10^{-19}$ & $1.50\times 10^{-16} \pm 4.34\times 10^{-19}$ &  &  \\
SN2007af  & 20070620 & $2.96\times 10^{-13} \pm 6.13\times 10^{-16}$ & $1.09\times 10^{-12} \pm 1.63\times 10^{-15}$ & 5762.2 & 6027.6 \\
SN2007af  & 20070625 & $2.15\times 10^{-13} \pm 5.45\times 10^{-16}$ & $7.80\times 10^{-13} \pm 2.19\times 10^{-15}$ &  &  \\
SN2007af  & 20070707 & $3.06\times 10^{-15} \pm 4.33\times 10^{-18}$ & $1.36\times 10^{-14} \pm 1.73\times 10^{-17}$ &  &  \\
SN2007af  & 20070710 & $8.47\times 10^{-14} \pm 4.97\times 10^{-16}$ & $3.08\times 10^{-13} \pm 1.39\times 10^{-15}$ &  &  \\
SN2007af  & 20070715 & $8.25\times 10^{-16} \pm 2.73\times 10^{-18}$ & $4.00\times 10^{-15} \pm 9.48\times 10^{-18}$ &  &  \\
SN2007af  & 20070718 & $7.49\times 10^{-14} \pm 6.87\times 10^{-16}$ & $3.47\times 10^{-13} \pm 3.23\times 10^{-15}$ &  &  \\
SN2007af  & 20070807 & $2.50\times 10^{-16} \pm 3.72\times 10^{-18}$ & $1.67\times 10^{-15} \pm 1.04\times 10^{-17}$ &  &  \\
SN2007af  & 20070821 & $3.57\times 10^{-16} \pm 4.16\times 10^{-18}$ & $2.15\times 10^{-15} \pm 1.93\times 10^{-17}$ &  &  \\
SN2007af  & 20080111 & $2.19\times 10^{-15} \pm 4.58\times 10^{-17}$ & $3.08\times 10^{-14} \pm 8.19\times 10^{-17}$ &  &  \\
SN2007gi  & 20080115 & $1.20\times 10^{-15} \pm 3.12\times 10^{-18}$ & $5.08\times 10^{-15} \pm 8.44\times 10^{-18}$ & 5706.8 & 6146.6 \\
SN2007le  & 20080827 & $1.80\times 10^{-17} \pm 4.41\times 10^{-19}$ & $2.02\times 10^{-16} \pm 5.68\times 10^{-19}$ & 5721.8 & 6087.5 \\
SN2007sr  & 20080623 & $4.42\times 10^{-14} \pm 5.84\times 10^{-17}$ & $3.05\times 10^{-13} \pm 4.04\times 10^{-16}$ & 5765.7 & 6059.7 \\
SN2008Q   & 20080828 & $9.20\times 10^{-17} \pm 2.08\times 10^{-19}$ & $6.25\times 10^{-16} \pm 5.91\times 10^{-19}$ & 5691.4 & 6089.5 \\
SN2009ig  & 20101017 & $1.50\times 10^{-15} \pm 6.87\times 10^{-18}$ & $1.23\times 10^{-14} \pm 1.85\times 10^{-17}$ & 5496.8 & 6156.7 \\
SN2009le  & 20101016 & $5.18\times 10^{-16} \pm 9.02\times 10^{-18}$ & $5.27\times 10^{-15} \pm 2.33\times 10^{-17}$ & 5587.7 & 6195.9 \\
SN2010ev  & 20101231 & $1.23\times 10^{-14} \pm 2.27\times 10^{-17}$ & $5.05\times 10^{-14} \pm 5.09\times 10^{-17}$ & 5649.8 & 6152.9 \\
SN2010ev  & 20110404 & $3.21\times 10^{-15} \pm 8.29\times 10^{-18}$ & $2.04\times 10^{-14} \pm 3.81\times 10^{-17}$ &  &  \\
SN2010gp  & 20110501 & $2.82\times 10^{-16} \pm 4.23\times 10^{-18}$ & $3.33\times 10^{-15} \pm 1.18\times 10^{-17}$ & 5726.6 & 6063.4 \\
SN2010hg  & 20110403 & $9.34\times 10^{-15} \pm 2.08\times 10^{-17}$ & $7.52\times 10^{-14} \pm 7.41\times 10^{-17}$ & 5728.8 & 6037.9 \\
SN2010kg  & 20110927 & $7.55\times 10^{-16} \pm 9.49\times 10^{-18}$ & $5.57\times 10^{-15} \pm 3.43\times 10^{-17}$ & 5650.3 & 6165.6 \\
SN2011ae  & 20120101 & $2.70\times 10^{-15} \pm 2.21\times 10^{-17}$ & $3.27\times 10^{-14} \pm 7.28\times 10^{-17}$ & 5684.3 & 6087.6 \\
SN2011at  & 20120227 & $1.93\times 10^{-15} \pm 8.90\times 10^{-18}$ & $1.32\times 10^{-14} \pm 2.58\times 10^{-17}$ & 5618.1 & 6118.7 \\
SN2011by  & 20111202 & $1.85\times 10^{-15} \pm 4.98\times 10^{-18}$ & $1.53\times 10^{-14} \pm 1.68\times 10^{-17}$ & 5723.6 & 6039.5 \\
SN2011by  & 20120315 & $5.84\times 10^{-17} \pm 4.32\times 10^{-19}$ & $8.72\times 10^{-16} \pm 1.26\times 10^{-18}$ &  &  \\
SN2011ek  & 20121011 & $1.73\times 10^{-16} \pm 4.75\times 10^{-18}$ & $1.12\times 10^{-15} \pm 9.34\times 10^{-18}$ & 5544.0 & 6130.4 \\
SN2011fe  & 20111123 & $7.06\times 10^{-12} \pm 1.78\times 10^{-15}$ & $1.76\times 10^{-11} \pm 5.31\times 10^{-15}$ & 5738.3 & 6027.2 \\
SN2011fe  & 20120102 & $1.38\times 10^{-12} \pm 2.86\times 10^{-16}$ & $5.87\times 10^{-12} \pm 1.76\times 10^{-15}$ &  &  \\
SN2011fe  & 20120324 & $2.14\times 10^{-13} \pm 5.88\times 10^{-17}$ & $1.81\times 10^{-12} \pm 5.28\times 10^{-16}$ &  &  \\
SN2011fe  & 20120427 & $8.30\times 10^{-14} \pm 2.34\times 10^{-17}$ & $8.94\times 10^{-13} \pm 2.50\times 10^{-16}$ &  &  \\
SN2011fe  & 20120612 & $5.69\times 10^{-14} \pm 1.77\times 10^{-17}$ & $8.42\times 10^{-13} \pm 2.62\times 10^{-16}$ &  &  \\
SN2011fe  & 20120720 & $2.83\times 10^{-14} \pm 4.42\times 10^{-17}$ & $5.26\times 10^{-13} \pm 3.84\times 10^{-16}$ &  &  \\
SN2011im  & 20121016 & $2.80\times 10^{-16} \pm 7.58\times 10^{-18}$ & $2.36\times 10^{-15} \pm 1.93\times 10^{-17}$ & 5687.0 & 6119.4 \\
SN2011iv  & 20121024 & $2.77\times 10^{-15} \pm 4.00\times 10^{-17}$ & $3.13\times 10^{-14} \pm 8.86\times 10^{-17}$ & 5693.8 & 6089.2 \\
SN2012cg  & 20130507 & $3.98\times 10^{-15} \pm 5.08\times 10^{-17}$ & $5.30\times 10^{-14} \pm 6.44\times 10^{-17}$ & 5698.9 & 6113.3 \\
SN2012cg  & 20130513 & $4.30\times 10^{-15} \pm 3.59\times 10^{-17}$ & $6.51\times 10^{-14} \pm 1.22\times 10^{-16}$ &  &  \\
SN2012cu  & 20130603 & $6.26\times 10^{-16} \pm 7.42\times 10^{-18}$ & $3.62\times 10^{-15} \pm 2.06\times 10^{-17}$ & 5692.6 & 6096.8 \\
SN2012fr  & 20130221 & $6.02\times 10^{-13} \pm 6.43\times 10^{-16}$ & $1.97\times 10^{-12} \pm 2.79\times 10^{-15}$ & 5759.2 & 6061.5 \\
SN2012fr  & 20130308 & $3.30\times 10^{-13} \pm 4.81\times 10^{-16}$ & $1.28\times 10^{-12} \pm 2.34\times 10^{-15}$ &  &  \\
SN2012fr  & 20130317 & $3.52\times 10^{-13} \pm 4.66\times 10^{-16}$ & $1.42\times 10^{-12} \pm 1.48\times 10^{-15}$ &  &  \\
SN2012fr  & 20130412 & $2.22\times 10^{-13} \pm 1.51\times 10^{-15}$ & $1.49\times 10^{-12} \pm 4.85\times 10^{-15}$ &  &  \\
SN2012fr  & 20130622 & $3.52\times 10^{-14} \pm 3.92\times 10^{-16}$ & $3.57\times 10^{-13} \pm 1.62\times 10^{-15}$ &  &  \\
SN2012fr  & 20130731 & $1.31\times 10^{-14} \pm 1.56\times 10^{-16}$ & $1.77\times 10^{-13} \pm 5.51\times 10^{-16}$ &  &  \\
SN2012fr  & 20131018 & $3.28\times 10^{-15} \pm 1.08\times 10^{-16}$ & $6.72\times 10^{-14} \pm 5.55\times 10^{-16}$ &  &  \\
SN2012fr  & 20131103 & $5.47\times 10^{-15} \pm 2.94\times 10^{-18}$ & $1.19\times 10^{-13} \pm 1.98\times 10^{-17}$ &  &  \\
SN2012fr  & 20131114 & $1.25\times 10^{-15} \pm 1.05\times 10^{-16}$ & $4.16\times 10^{-14} \pm 3.91\times 10^{-16}$ &  &  \\
SN2012hr  & 20131006 & $3.61\times 10^{-17} \pm 2.28\times 10^{-19}$ & $5.34\times 10^{-16} \pm 7.39\times 10^{-19}$ & 5775.3 & 6041.1 \\
SN2012hr  & 20131230 & $2.53\times 10^{-16} \pm 2.76\times 10^{-17}$ & $7.26\times 10^{-15} \pm 1.39\times 10^{-16}$ &  &  \\
SN2013aa  & 20130710 & $3.75\times 10^{-13} \pm 4.69\times 10^{-16}$ & $1.60\times 10^{-12} \pm 1.07\times 10^{-15}$ & 5720.3 & 6055.9 \\
SN2013aa  & 20130827 & $5.88\times 10^{-14} \pm 1.19\times 10^{-16}$ & $4.14\times 10^{-13} \pm 5.40\times 10^{-16}$ &  &  \\
SN2013aa  & 20130913 & $6.54\times 10^{-14} \pm 4.35\times 10^{-16}$ & $5.13\times 10^{-13} \pm 1.62\times 10^{-15}$ &  &  \\
SN2013aa  & 20140131 & $1.76\times 10^{-15} \pm 1.48\times 10^{-16}$ & $5.26\times 10^{-14} \pm 3.92\times 10^{-16}$ &  &  \\
SN2013aa  & 20140216 & $2.10\times 10^{-15} \pm 2.73\times 10^{-17}$ & $3.78\times 10^{-14} \pm 4.23\times 10^{-17}$ &  &  \\
SN2013aa  & 20140422 & $8.32\times 10^{-16} \pm 2.73\times 10^{-17}$ & $1.61\times 10^{-14} \pm 3.68\times 10^{-17}$ &  &  \\
SN2013cs  & 20140322 & $3.28\times 10^{-16} \pm 9.26\times 10^{-17}$ & $7.49\times 10^{-15} \pm 3.94\times 10^{-16}$ & 5735.8 & 6061.7 \\
SN2013cs  & 20140324 & $9.99\times 10^{-16} \pm 1.80\times 10^{-17}$ & $1.56\times 10^{-14} \pm 2.24\times 10^{-17}$ &  &  \\
SN2013ct  & 20131218 & $3.71\times 10^{-14} \pm 8.54\times 10^{-17}$ & $3.35\times 10^{-13} \pm 2.32\times 10^{-16}$ & 5753.1 & 6030.3 \\
SN2013dl  & 20141130 & $1.17\times 10^{-14} \pm 6.49\times 10^{-17}$ & $5.87\times 10^{-14} \pm 1.04\times 10^{-16}$ & 5763.5 & 6061.3 \\
SN2013dl  & 20141224 & $7.29\times 10^{-15} \pm 6.95\times 10^{-17}$ & $4.47\times 10^{-14} \pm 8.46\times 10^{-17}$ &  &  \\
SN2013dy  & 20140626 & $5.54\times 10^{-15} \pm 3.61\times 10^{-17}$ & $6.29\times 10^{-14} \pm 3.65\times 10^{-15}$ & 5636.9 & 6168.7 \\
SN2013dy  & 20140920 & $1.45\times 10^{-14} \pm 1.34\times 10^{-16}$ & $1.00\times 10^{-13} \pm 6.47\times 10^{-15}$ &  &  \\
SN2013ef  & 20141224 & $3.28\times 10^{-15} \pm 3.58\times 10^{-17}$ & $1.51\times 10^{-14} \pm 5.48\times 10^{-17}$ & 5733.8 & 6053.0 \\
SN2013gy  & 20140920 & $7.63\times 10^{-15} \pm 5.97\times 10^{-17}$ & $5.72\times 10^{-14} \pm 1.82\times 10^{-15}$ & 5726.8 & 6064.7 \\
SN2014J   & 20140920 & $1.65\times 10^{-12} \pm 7.92\times 10^{-15}$ & $5.19\times 10^{-12} \pm 9.56\times 10^{-15}$ & 5737.8 & 6069.5 \\
\enddata
\end{deluxetable}

\end{document}